# Beyond the Petermann limit: can exceptional points increase sensor precision?

DAVID D. SMITH[1,*], HONGROK CHANG[2], EUGENIY MIKHAILOV[3], AND SELIM M. SHAHRIAR[4,5]

[1]NASA Marshall Space Flight Center, Space Systems Department, ES23, Huntsville, AL 35812, USA
[2]Center for Applied Optics, University of Alabama in Huntsville, Huntsville, AL 35899, USA
[3]Department of Physics, College of William & Mary, Williamsburg, Virginia 23187, USA
[4]Northwestern University, Department of Electrical and Computer Engineering, Evanston, IL 60208, USA
[5]Northwestern University, Department of Physics and Astronomy, Evanston, IL 60208, USA
*david.d.smith@nasa.gov

Experiments near the lock-in region in maximally dissipative non-Hermitian systems, e.g., conventional laser gyroscopes near the deadband, have run up against the Petermann limit, where excess noise exactly cancels any scale-factor enhancement resulting in no overall enhancement in precision. As a result, one might be tempted to conclude that exceptional points (EPs) generally cannot be used to increase the precision of laser sensors. Indeed, using a linear eigenmode analysis we show that the Petermann limit applies not just to maximally dissipative systems, but for any type of EP, owing to the fact that EPs are rotationally invariant. It turns out, however, that this restriction comes from the assumption of linearity. We find that nonlinearity breaks the rotation symmetry such that the different types of EPs are no longer equivalent above threshold. In particular, for sufficiently high measurement frequencies, EPs in conservatively coupled systems having a saturation imbalance can lead to an increase in the fundamental precision beyond the Petermann limit. Importantly, we find that only one mode lases under these conditions. We show that the beat note can be recovered by interference with an auxiliary mode, but that this has consequences for the quantum and classical noise that depend on the recovery scheme. Thus, it remains to be seen whether practical experiments can be designed that can take advantage of this enhancement.

## I. INTRODUCTION

Exceptional points (EPs) are singularities that arise in non-Hermitian systems where the Hamiltonian becomes defective, the eigenstates become maximally non-orthogonal, and the eigenvalues and eigenvectors simultaneously coalesce into a degeneracy. Prominent examples of these non-conservative systems in the field of optics include coupled resonators (CRs) and ring laser gyroscopes (RLGs) [1-5], materials with periodic potentials [6-9], and systems with parity-time (PT) symmetry [10-15]. One reason EPs have been of recent interest is because the sensitivity of the frequency difference between the eigenstates to an external perturbation, i.e., the scale factor, has been shown to diverge at an EP [16-22]. The boost in scale-factor sensitivity has now been demonstrated experimentally in passive and active fast-light cavities [21-27], optomechanical and nanoparticle detection schemes [30, 31], and CRs including RLGs [32-37].

    A sensitivity enhancement, by itself, is not sufficient to enhance measurement precision, however. In particular, the enhancement in scale factor may lead to a concomitant increase in measurement uncertainty. Indeed, it has long been established that in non-Hermitian laser systems the fundamental laser linewidth increases above the Schawlow-Townes limit as a result of the non-orthogonality of the resonant modes. This broadening is characterized (at least in the linear limit) by the Petermann excess-noise factor, which has been shown to diverge at an EP [38-45]. The excess noise arises from correlations that occur between the noise sources of different eigenmodes when they are not orthogonal. In effect, the spontaneous emission in a given mode is affected not only by noise photons in the same mode, but also by those in the

other modes. The divergence in linewidth occurs because the EP is where the eigenmodes become maximally non-orthogonal [41]. In this light, the crucial question for sensing applications is whether the sensitivity diverges faster than the linewidth as the singularity is approached.

We have pointed out previously that there are different types of EPs in CRs [46]. One type of EP forms when the coupling is conservative, there is no detuning, and the loss difference is balanced by the coupling. This EP is also parity-time (PT) symmetric. A second type of EP occurs when the coupling is maximally dissipative, there is no loss difference, and the detuning is balanced by the coupling. This EP is not PT-symmetric and corresponds to the edge of a lock-in region, such as the deadband that occurs in conventional RLGs. Yet a third type of EP is partially dissipative, involving a mixture of the two types of coupling above. An important question is whether the divergences of the sensitivity and linewidth are the same for all these different types of EPs. If so, a straightforward way to determine whether precision can be enhanced would be to measure these quantities near the RLG deadband edge. In fact, such an experiment has recently been performed [35], which found that the excess noise completely counteracted the increase in sensitivity, resulting in no overall benefit to measurement precision near this maximally dissipative EP. In this work we attempt to answer whether that result is specific to the particular type of EP and the specific parameters used, or holds more generally for any type of EP.

We start with a general linear eigenmode approach to determine the enhancement in precision for any type of EP and show that there is no set of parameters that results in an enhancement in this linear limit. Furthermore, we demonstrate that changing from one type of EP to another is equivalent to a simple rotation in parameter space, so measurements performed at any one type of EP also apply to other types. (Sec. II-IV). We then turn our attention to approaches that account for the nonlinearity that is characteristic of lasers pumped above threshold. We find that the nonlinearity breaks the rotation symmetry such that the different types of EPs are no longer equivalent. As a consequence, in contrast with the linear results, we find that an enhancement in the quantum-limited precision can occur when the coupling is conservative (Sec. V-VIII). In Sec. V a quasilinear approach is taken where the gain coefficients in the Petermann factor are simply replaced by their saturated steady-state counterparts, whereas in section VI the nonlinear coupled equations are linearized about the steady-state to obtain the noise spectrum. By taking this more rigorous approach we find deviations from the frequency-independent quasilinear Petermann factor due to the coloring of the noise, the coupling of the noise fluctuations, and the fact that the threshold conditions for the coupled and uncoupled systems are different. Furthermore, in Sec. VII, we show that because the coloring is described by different characteristic decay rates for the two systems, an enhancement in precision is possible. In Sec. VIII, we show that only one mode lases in these systems, resulting in zero beat frequency. Unlike the situation within the gyro deadband, however, the beat frequency can be recovered because there is a frequency shift for the remaining lasing mode. We describe a simple experimental approach that demonstrates the beat frequency recovery by interference with an auxiliary mode, but show that any such scheme will incur consequences for the quantum and classical noise. Finally, in Sec. IX we show how our work explains the results of previous experiments in these non-Hermitian systems.

## II. SCALE-FACTOR ENHANCEMENT

Consider the problem of two CRs as shown in Fig. 1. The resonators have resonant frequencies $\omega_1$ and $\omega_2$ when they are uncoupled from one another, and any additional losses not due to the coupling between the resonators are represented by the photon loss rates $\gamma_1$ and $\gamma_2$.

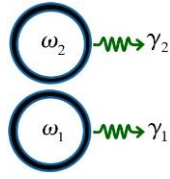

FIG. 1. Two coupled resonators.

The coupled-mode equations describing the field dynamics are:

$$\dot{E}_1(t) = -i\tilde{\omega}_1 E_1(t) + i\frac{\tilde{\kappa}_1}{2} E_2(t)$$
$$\dot{E}_2(t) = -i\tilde{\omega}_2 E_2(t) + i\frac{\tilde{\kappa}_2}{2} E_1(t) \quad (1)$$

where $\tilde{\omega}_i = \omega_i - i\gamma_i/2$ $(i=1,2)$ are complex frequencies, and $\tilde{\kappa}_i$ are complex coupling coefficients. We define a complex detuning between the resonators as $\tilde{\delta} \equiv \tilde{\omega}_1 - \tilde{\omega}_2 = \delta - i\gamma/2$, where $\delta = \omega_1 - \omega_2$ is the real-valued detuning and $\gamma = \gamma_1 - \gamma_2$ is the difference in the loss rates. The eigenvalues of this system of equations are

$$\tilde{\omega}_\pm = \tilde{\omega}_{avg} \pm \frac{\tilde{\Omega}}{2} = \omega_\pm - i\frac{\gamma_\pm}{2}, \quad (2)$$

where $\tilde{\Omega} = (\tilde{\delta}^2 + \tilde{\kappa}_1\tilde{\kappa}_2)^{1/2}$ is the generalized Rabi frequency, and $\tilde{\omega}_{avg} = (\tilde{\omega}_1 + \tilde{\omega}_2)/2 = \omega_{avg} - i\gamma_{avg}/2$. The eigenvalues are in general complex, but can be decomposed into real-valued frequencies $\omega_\pm$ and linewidths $\gamma_\pm$, as shown.

Rotating to slowly varying amplitudes and introducing Langevin noise terms $f_i$ results in

$$\frac{d}{dt}\begin{pmatrix} e_1 \\ e_2 \end{pmatrix} = \frac{i}{2}\begin{pmatrix} -\delta + i\gamma_1 & \tilde{\kappa}_1 \\ \tilde{\kappa}_2 & \delta + i\gamma_2 \end{pmatrix}\begin{pmatrix} e_1 \\ e_2 \end{pmatrix} + \begin{pmatrix} f_1 \\ f_2 \end{pmatrix}. \quad (3)$$

This unitary transformation eliminates $\omega_{avg}$ from Eq. (2). The noise terms have correlations $\langle f_i(t)^* f_j(t')\rangle = R_i^{sp}\delta_{ij}\delta(t-t')$ where $R_i^{sp}$ is the spontaneous emission rate in each resonator. Note that while $f_1$ and $f_2$ are uncorrelated, a non-unitary transformation is required to diagonalize the matrix above and this correlates the noise sources of the resultant eigenmodes, leading to Petermann excess noise (see section III).

For convenience we will assume $\tilde{\kappa}_1 = \tilde{\kappa}_2 = \tilde{\kappa}$ such that the Hamiltonian is symmetric. In this case, the generalized Rabi frequency is $\tilde{\Omega} = (\tilde{\delta}^2 + \tilde{\kappa}^2)^{1/2}$, i.e.,

$$\tilde{\Omega} = \left[\delta^2 - (\kappa'')^2 + (\kappa')^2 - (\gamma/2)^2 + i(2\kappa'\kappa'' - \delta\gamma)\right]^{1/2}, \quad (4)$$

where $\tilde{\kappa} \equiv \kappa\exp(i\theta)$, with $\kappa = [(\kappa')^2 + (\kappa'')^2]^{1/2}$ and $\theta = \operatorname{atan}(\kappa''/\kappa')$. An EP occurs when the eigenvalues are fully degenerate in both frequency and linewidth, i.e., when $\tilde{\Omega} = 0$. There are three cases to consider: (*i*) **conservative coupling** ($\tilde{\kappa} = \kappa'$) An EP occurs when $\delta = 0$, and $|\gamma/2| = |\kappa'|$. This EP corresponds to the parity-time (PT) symmetric phase transition; (*ii*) **maximally dissipative coupling** ($\tilde{\kappa} = i\kappa''$). An EP occurs when $\gamma = 0$ and $|\delta| = |\kappa''|$. This EP corresponds to the edge of the deadband, the region of zero sensitivity that occurs in conventional RLGs [2]; and (*iii*) **partially dissipative coupling** ($\tilde{\kappa} = \kappa' + i\kappa''$). An EP forms when $\gamma \neq 0$, $|\kappa'| = |\gamma/2|$, $|\delta| = |\kappa''|$, and $\kappa'\kappa'' = \delta\gamma/2$. These EPs are located along the unit circle shown in Fig. 2. Other EP locations are possible if one relaxes the assumption that $\tilde{\kappa}_1$ and $\tilde{\kappa}_2$ are equal in magnitude [37].

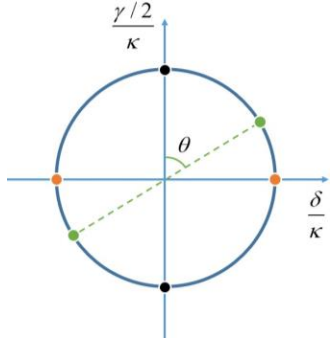

FIG. 2. Parameters values for EPs. Conservative, maximally dissipative, and partially dissipative EPs are represented on the unit circle by black, orange and green dots, respectively. Zero-sensitivity bands occur along the diameter between any pair of EPs of the same color.

We can rewrite Eq. (4) as $\tilde{\Omega} = (A + iB)^{1/2}$. The magnitude of the scale-factor enhancement is then [46]

$$|S| \equiv \left|\frac{d\Omega'}{d\delta}\right| = \left|\frac{\tilde{\delta}}{2(\tilde{\delta}^2 + \tilde{\kappa}^2)^{1/2}} + c.c.\right|, \tag{5}$$

where

$$\Omega' = \omega_+ - \omega_- = \frac{1}{\sqrt{2}}\left[A + \sqrt{A^2 + B^2}\right]^{1/2} \tag{6}$$

represents the relative detuning between the real parts of the eigenvalues or beat frequency [47]. For case (i) $A = \delta^2 + (\kappa')^2 - (\gamma/2)^2$ and $B = \delta\gamma$. At sufficiently small detunings $|\delta|$ we obtain [46, 48]:

$$|S| = \begin{cases} |\gamma/2|/[(\gamma/2)^2 - (\kappa')^2]^{1/2} & |\kappa'| < |\gamma/2| \\ |\gamma/(2\delta)|^{1/2}/2 & |\kappa'| = |\gamma/2| \\ |\delta|(\kappa')^2/[(\kappa')^2 - (\gamma/2)^2]^{3/2} & |\kappa'| > |\gamma/2| \end{cases}, \tag{7}$$

where $\delta^2 \ll \gamma^2$ applies to the middle equation, and $\delta^2 \ll |(\kappa')^2 - (\gamma/2)^2|$ applies to the other two equations. Note that when $\delta = 0$ the middle equation predicts a divergence in the sensitivity at the EP, whereas the last equation predicts the sensitivity drops to zero over a range of $\gamma$ values corresponding to the unbroken (exact) PT-symmetry region (between the black dots) in Fig. 2.

For case (ii) the beat frequency is also found from Eq. (6) with $B = \delta\gamma$, but now $A = \delta^2 - (\kappa'')^2 - (\gamma/2)^2$. Thus, the relations for the scale-factor enhancements and their associated regimes of validity can be found by simply interchanging $\delta$ and $\gamma/2$, and replacing $\kappa'$ with $\kappa''$ in Eq. (7) [46]. The resulting equations apply for sufficiently small loss differences $|\gamma|$. In this case the zero-sensitivity region occurs over a range of detunings corresponding to the deadband region (between the orange dots) in Fig. 2. Note that the appearance of these zero-sensitivity regions implies that, provided the linewidth is nonzero, the precision will drop precipitously to zero in these regions.

### III. PETERMANN EXCESS-NOISE FACTOR

The Petermann excess-noise factor can be obtained from [42, 49]

$$K = \frac{1}{1-|\langle e_+|e_-\rangle|^2}, \tag{8}$$

where the instantaneous eigenmodes of Eq. (1) are given by

$$|e_\pm\rangle = N_\pm \begin{pmatrix} \tilde{\kappa} \\ \tilde{\delta} \mp \tilde{\Omega} \end{pmatrix}, \tag{9}$$

and $N_\pm = (|\tilde{\kappa}|^2 + |\tilde{\delta} \mp \tilde{\Omega}|^2)^{-1/2}$ are normalization factors. The eigenmodes are not in general orthogonal but are skewed such that $\langle e_\pm|e_\mp\rangle \neq 0$. They are, however, biorthogonal, i.e., $\langle e_\pm^*|e_\mp\rangle = 0$. Our choice of normalization is such that $\langle e_\pm|e_\pm\rangle = 1$, which governs the form of Eq. (8).

For case (*i*), the excess noise factor is

$$K = \begin{cases} (\gamma/2)^2/[(\gamma/2)^2 - (\kappa')^2] & |\kappa'|<|\gamma/2| \\ |\gamma/(2\delta)|/2 & |\kappa'|=|\gamma/2| \\ (\kappa')^2/[(\kappa')^2 - (\gamma/2)^2] & |\kappa'|>|\gamma/2| \end{cases}, \tag{10}$$

where the middle equation applies for small detunings, i.e., $\delta^2 \ll \gamma^2$, but the other two equations are derived at $\delta = 0$ with no approximations and are the same as those found in [42]. Again, for case (*ii*) we simply interchange $\delta$ and $\gamma/2$, and replace $\kappa'$ with $\kappa''$ in Eq. (10).

For sensors utilizing resonant optical cavities the error in the determination of the measurand (such as the rotation rate for an RLG) scales as $\sigma/s$, where $\sigma$ is the uncertainty to which the center of the resonance line can be determined and $s$ is the scale-factor which relates changes in the resonance frequency to changes in the measurand [25]. When the resonant cavity is a laser operating under quantum-limited conditions in the white frequency noise limit, $\sigma$ is proportional to the square-root of the Schawlow-Townes linewidth [50, 51]. The enhancement in precision in this case is, therefore, $|S|/K^{1/2}$. Comparing Eq. (7) and Eq. (10) we obtain

$$\frac{|S|}{K^{1/2}} = \begin{cases} 1 & |\kappa'|<|\gamma/2| \\ (1/2)^{1/2} & |\kappa'|=|\gamma/2| \\ 0 & |\kappa'|>|\gamma/2| \end{cases}, \tag{11}$$

which applies at $\delta = 0$. Similar relations apply for case (*ii*) at $\gamma/2 = 0$ after interchanging the parameters. These results show that the best precision is obtained in the broken PT-symmetry regime (or outside the deadband for case (*ii*)), but that the precision drops precipitously to zero in the PT-symmetric region (or inside the deadband), and that the EP represents a transition between these two sets of behavior. In Fig. 3 the value of $|S|/K^{1/2}$, obtained by evaluating Eq. (8) and the RHS of Eq. (5), is plotted versus the detuning and the loss difference. Note that the values of the curves at $\delta = 0$ in Fig. 3(a), as well as at the points indicated on the dashed line in Fig. 3(b), correspond to those found in Eq. (11).

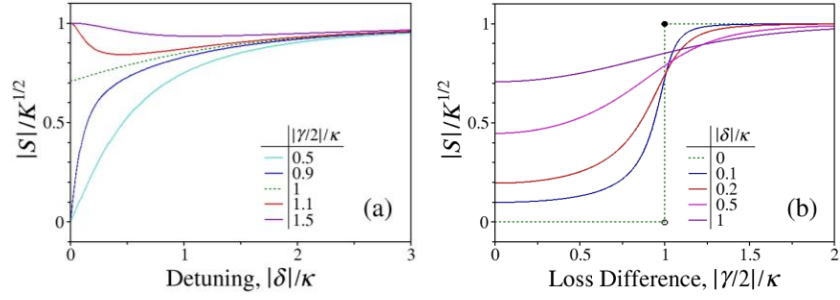

FIG. 3. $|S|/K^{1/2}$ plotted vs. (a) detuning and (b) loss difference for case (i). In (a) the bottom two curves are PT-symmetric at $\delta = 0$ and approach $|S|/K^{1/2} = 0$, whereas the top two curves are in the PT-symmetry broken regime, and approach $|S|/K^{1/2} = 1$. All the curves approach $|S|/K^{1/2} = 1$ at large at $|\delta|$. In (b) a discontinuity is observed between the broken and unbroken PT-symmetry regimes, showing the effect of the zero-sensitivity region. The maximum and minimum values at the EP are indicated by solid and open black dots, respectively.

In Fig. 4, the enhancement in precision $|S|/K^{1/2}$ is plotted vs. the detuning and the loss difference. The value of $|S|/K^{1/2}$ never rises above unity. Indeed, a deep dipole-shaped hole is formed between the EPs, demonstrating the deleterious effect of the EPs. Both Fig. 3 and Fig. 4 are plotted for conservative coupling (case (i)), but the plots for the maximally dissipative result (case (ii)) are identical as they are obtained by a simple interchange of $\delta$ and $\gamma/2$, i.e., a rotation of $\theta = 90°$ in Figs. 2 and 4. The partially dissipative case is obtained by a rotation of $\theta = \operatorname{atan}(\kappa''/\kappa')$. In this case $\delta$ and $\gamma/2$ must be varied simultaneously to reproduce the results shown in Figs. 3(a) and 3(b), because the zero-sensitivity region now occurs along the green dashed line in Fig. 2.

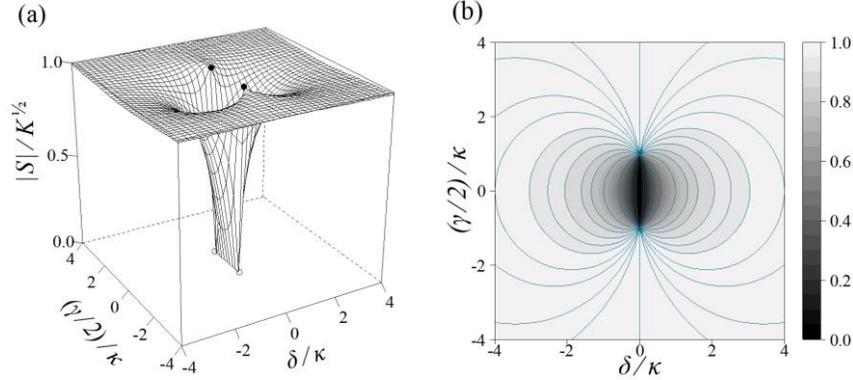

FIG. 4. A hole forms in the enhancement in precision in the vicinity of the EPs. In (a) maximum and minimum values at the EP are indicated by solid and open black dots, respectively. The contour plot in (b) reveals dipole-like curves of constant $|S|/K^{1/2}$, with the hole becoming circular when $|S|/K^{1/2} = (1/2)^{1/2}$.

This brings up an important point: EPs are invariant under rotation in the parameter space spanned by the detuning and the loss difference. In practice there is a difference between EP types because a single tuning parameter is typically used. For example, if $\delta$ is the tuning parameter then for maximally dissipative coupling $|S|/K^{1/2}$ would be unity outside the deadband (Fig. 3(b)), whereas for conservative coupling $|S|/K^{1/2}$ would drop below unity as the system is detuned (Fig. 3(a)). Nevertheless, we take advantage of this invariance in the next section to prove that $|S|/K^{1/2} \leq 1$ for *any* choice of parameters.

## IV. GENERAL LINEAR SOLUTION

Note that Eq. (11) only applies at $\delta = 0$ for conservative coupling and at $\gamma/2 = 0$ for maximally dissipative coupling, and so does not predict the value of $|S|/K^{1/2}$ at the other parameter values shown in Fig. 3. In addition, we have yet to derive a relation for the case of partially dissipative coupling. To deal with this more general situation, first note that the excess-noise factor can be obtained directly from the Hamiltonian $H$ representing Eq. (1) through the relations [52]

$$H_0 = H - Tr(H)/2 = \begin{pmatrix} -\tilde{\delta} & \tilde{\kappa} \\ \tilde{\kappa} & \tilde{\delta} \end{pmatrix}, \tag{12}$$

and

$$K = \frac{1}{2}\left[1 + \frac{Tr(H_0^\dagger H_0)}{|Tr(H_0^2)|}\right] = \frac{1}{2}\left[1 + \frac{|\tilde{\delta}|^2 + \kappa^2}{|\tilde{\Omega}|^2}\right]. \tag{13}$$

The relation for $|S|$ is then obtained by setting $A = (\kappa')^2 - (\gamma/2)^2 + \delta^2 - (\kappa'')^2$ and $B = 2\kappa'\kappa'' - \delta\gamma$ in Eq. (6) and differentiating to obtain

$$\frac{|S|}{K} = \left|\frac{\delta}{\Omega'}\right|\left\{1 - \frac{\kappa''[\kappa'' + \kappa'\gamma/(2\delta)]}{(\Omega')^2 + (\gamma/2)^2 + (\kappa'')^2}\right\}, \tag{14}$$

where the relations $(A^2 + B^2)^{1/2} = |\tilde{\Omega}|^2$, and $|\tilde{\delta}|^2 + \kappa^2 + |\tilde{\Omega}|^2 = 2[(\Omega')^2 + (\gamma/2)^2 + (\kappa'')^2]$ have been used. By taking the limit $\delta \to \kappa''$ and administering L'hospital's rule we then find that $|S|/K^{1/2} = \sqrt{1/2}$ for any type of EP.

Indeed, with the general solution in hand, it is now straightforward to show that $|S|/K^{1/2} \leq 1$ for any set of parameters. First, for conservative coupling we know from Eq. (14) that $|S|/K = |\delta/\Omega'|$, where $\Omega' = \text{Re}\left\{\sqrt{(\delta - i\gamma/2)^2 + (\kappa')^2}\right\}$. We can then recast Eq. (14) by rotating the plane defined by the variables $\delta$ and $\gamma/2$ through the angle $\theta = \text{atan}(\kappa''/\kappa')$ shown in Fig. 2, i.e.,

$$\frac{|S|}{K} = \left|\frac{\delta_R}{\Omega'_R}\right|, \tag{15}$$

where $\Omega'_R = \text{Re}\left\{\sqrt{(\delta_R - i\gamma_R/2)^2 + \kappa^2}\right\}$ [53] and

$$\begin{pmatrix} \delta_R \\ \gamma_R/2 \end{pmatrix} = \begin{bmatrix} \cos\theta & -\sin\theta \\ \sin\theta & \cos\theta \end{bmatrix}\begin{pmatrix} \delta \\ \gamma/2 \end{pmatrix}. \tag{16}$$

Now, from Eq. (6) it is straightforward to show that for conservative coupling $|\Omega'| \geq |\delta|$ for all values of $\delta$. Furthermore, because the general case is simply a rotation of the solution for conservative coupling, it then follows that $|S|/K \leq 1$ for any choice of parameters.

In fact a tighter bound can be established because $|\Omega'|$ is also always greater than or equal to $|S\delta|$ for conservative coupling. To prove that this is the case is equivalent to demonstrating that $K \leq |\Omega'/\delta|^2$, which by substitution of Eq. (13) corresponds to the quadratic equation

$$|\tilde{\Omega}|^4 + |\tilde{\Omega}|^2\left[(\kappa')^2 - (\gamma/2)^2\right] - \delta^2\left[\delta^2 + (\kappa')^2 + (\gamma/2)^2\right] \geq 0. \tag{17}$$

This equation cannot have more than one real root, and is therefore satisfied for any choice of $\delta$, $\gamma$, and $\kappa'$. Again, transforming to the rotated coordinates as described above, we find that $|S|/K^{1/2} \leq 1$ for any choice of parameters, with the maximum value of unity obtained near to and far away from the EP as shown in Fig. 4.

## V. ABOVE THRESHOLD BEHAVIOR: QUASILINEAR APPROACH

Until now our eigenmode analysis has been entirely linear, i.e., we've assumed the parameters $\delta, \gamma, \kappa',$ and $\kappa''$ do not depend on the fields. Therefore, in a strict sense the results do not apply above threshold, where $\gamma$ depends on the intracavity intensities as a result of gain (and possibly absorption) saturation. Nevertheless, if one is only interested in steady-state behavior, the linear theory has been shown to work well in some cases for systems pumped above threshold [35, 36, 37, 46] provided saturated (steady-state) values of the gain coefficients are used in the analysis. In this quasilinear approach, the beat frequency and the excess-noise factor are still determined by Eq. (6) and (13), respectively (with substitution of saturated gain values). On the other hand, the scale-factor enhancement must be modified owing to the dependence of $\gamma$ on the detuning, $\Omega'(\delta, \gamma(\delta))$, as dictated by the threshold condition [46]. In accordance with the chain rule,

$$S = \frac{d\Omega'}{d\delta} = \frac{\partial \Omega'}{\partial \delta} + \frac{\partial \Omega'}{\partial \gamma}\frac{\partial \gamma}{\partial \delta} \ . \tag{18}$$

Note that from the first term on the R.H.S. we recover Eq. (14) after dividing by $K$. Thus, if not for the additional second term, our previous conclusion that $S^2/K \leq 1$ would still stand. The extra term, however, leads to situations where the scale factor diverges faster than expected [37, 46, 54] such that $S^2/K$ can increase above unity. After dividing by $K$, the second term can be written as

$$\left|\frac{\delta}{\Omega'}\right|\left|\frac{(\Omega')^2 - \delta^2 + 2\kappa'\kappa''\delta/\gamma}{(\Omega')^2 + (\gamma/2)^2 + (\kappa'')^2}\right|\left|\frac{\gamma}{2\delta}\right|\psi \xrightarrow{EP} \frac{1}{|2S|}\left|\frac{\gamma}{2\delta}\right|\psi_{EP}, \tag{19}$$

where the factor $\psi = |\partial_\delta \gamma / 2|$ represents the saturation imbalance. Just as before, we've taken the limit $\delta \to \kappa''$ and administered L'hospital's rule to obtain the result on the R.H.S. of Eq. (19), which applies at the EP. Pulling together the results at the EP from Eqs. (14) and (19) we have

$$\frac{S^2}{K} = \frac{1}{2}\left(1 + \left|\frac{\kappa'}{\kappa''}\right|\psi_{EP}\right), \tag{20}$$

where the factor $\psi_{EP}$ should be evaluated at the EP after taking the partial derivative. The term outside the parentheses is the result without saturation, and is also obtained when $\psi = 0$ because the saturation is balanced, resulting in no change in $S$ or $K$. On the other hand, $S^2/K$ increases above unity when

$$|\kappa'|\psi_{EP} > |\kappa''|. \tag{21}$$

To evaluate the partial derivative $\psi$ in the steady state, we set $\gamma_\pm = \gamma_{avg} \mp \Omega'' = 0$, and use the fact that $(\Omega'')^2 = (\Omega')^2 - A$, to obtain the threshold condition $A_0 = (B_0/2\gamma_{avg,0})^2 - \gamma_{avg,0}^2$, which for conservative coupling becomes

$$(\kappa')^2 = -\gamma_{1,0}\gamma_{2,0}[1 + (\delta/\gamma_{avg,0})^2], \tag{22}$$

where the subscript '0' indicates steady state. Restricting the result to the steady state in this way essentially collapses the plot in Fig. 4(a) to two dimensions because not all the values along the $\gamma$-axis are accessible. This allows a determination of $S^2/K$ in the steady state when the system is pumped above threshold, either directly at the EP via Eq. (20), or in the more general case of arbitrary parameters by using Eqs. (18), (19), and (14). Unfortunately, except for the particular case of $\psi = 0$ (balanced saturation), Eq. (22) does not uniquely determine $\psi_0$ if both $\gamma_1$ and $\gamma_2$ are saturating, i.e., it requires one of these coefficients to be held constant. When both resonators saturate we must instead resort to numerical solution of the coupled nonlinear equations that result from replacing the constant coefficients $\gamma_1$ and $\gamma_2$ in Eq. (1) with saturable ones of the form $\gamma_i(I_i) = \gamma_i^s(I_i) + \gamma_i^n = \gamma_i^u/(1+\beta_i I_i) + \gamma_i^n$ $(i=1,2)$ where $\beta_i$ are the self-saturation coefficients, $I_i$ are the intensities in the resonators, and $\gamma_i^s$, $\gamma_i^n$, and $\gamma_i^u$ are the saturable, nonsaturable (constant), and unsaturated loss coefficients, respectively [55].

Eq. (20) shows that the greatest enhancement in $S^2/K$ is achieved at the conservative coupling EP where $\kappa'' = 0$. In Fig. 5 the beat frequency $\Omega'$ (Fig. 5a) as well as the value of $|S|/K^{1/2}$ (Fig. 5b) are plotted vs. $\delta$ for conservative coupling at increasing values of the saturation imbalance $\psi_0$, with the system set to saturate to the EP at threshold in each case. Three cases are shown: (*i*) the gain in both resonators saturates equally (blue curve), but the nonsaturable loss coefficients are different, i.e., $\gamma_1^n/\kappa' = 0$ and $\gamma_2^n/\kappa' = 2$; (*ii*) only the first resonator saturates (green curve) with equal nonsaturable losses $\gamma_1^n/\kappa' = \gamma_2^n/\kappa' = 1$. (*iii*) the saturation is equal in magnitude but opposite in sign (red curve), i.e., the gain (loss) in the first (second) resonator saturates. The nonsaturable losses are both set to zero in this case, i.e., $\gamma_1^n/\kappa' = \gamma_2^n/\kappa' = 0$. For each case we've set $\beta_1 = \beta_2 = 1$. The saturation imbalance is $\psi_0 = 0$, $\psi_0 = |\partial_\delta \gamma_{1,0}|$, and $\psi_0 = 2|\partial_\delta \gamma_{1,0}|$, respectively. In Fig. 5(a), the yellow surface represents the beat frequency obtained from Eq. (6) for all possible combinations of $\delta$ and $\gamma$. The curves, on the other hand, result from restricting the result to steady-state. The two-dimensional projections of these curves are shown in the inset of Fig. 5(b). The beat frequency changes nonlinearly near the EP at $\delta = 0$ in each case. The data points in Fig. 5(b) are from numerical solutions of the nonlinear coupled equations [56] for the saturated gain coefficients, which were then substituted into Eqs. (6) and (13) to obtain $\Omega'$ and $K$, respectively. The scale factor enhancement was then calculated numerically from $\Omega'$. On the other hand, the solid curves (with the exception of the red curve), are analytic solutions of Eq. (18) with $\psi_0$ determined from Eq. (22). The good agreement validates these relations. For the red solid curve both resonators saturate, and do so at different rates with increasing $\delta$, so the steady state value of $\psi$ must be determined from the numerical computation.

The effect of the saturation imbalance is clear: when $\psi_0 = 0$ the value of $\gamma_0$ is independent of $\delta$, resulting in no change from the linear result, but for nonzero $\psi_0$ the curves in Fig. 5a take trajectories of varying $\gamma_0$, leading to an increase in $S$ and thus in $|S|/K^{1/2}$ as shown in Fig. 5(b). For case (*i*) $\psi_0 = 0$ and the gain saturation has no effect. The beat frequency displays the same square-root relation predicted by the linear equations, and $|S|/K^{1/2}$ is the same as the dashed curve in Fig. 3(a), remaining below unity for all detunings. For case (*ii*), on the other hand, the saturation imbalance $\psi_0 = |\partial_\delta \gamma_{1,0}|$ causes the scale factor to increase more than expected, and $|S|/K^{1/2}$ exceeds unity over a large bandwidth of about 0.2 in dimensionless units. For case (*iii*) $\psi_0 = 2|\partial_\delta \gamma_{1,0}|$, which maximizes the saturation imbalance. Aside from the narrow cusp around $\delta = 0$, the response is the same as that of the PT-symmetric case in the absence of gain saturation with $\kappa' = 2|\gamma_0/2|$. The enhancement bandwidth effectively drops to zero in this case. Thus, there is an inverse relationship between the bandwidth and the maximum enhancement. In addition, according to Eq. (21), for pure conservative coupling

($\kappa'' = 0$) the enhancement always diverges unless the saturation is balanced. This absurdity is obviated by the fact that in practice there will always be some dissipation involved in the coupling.

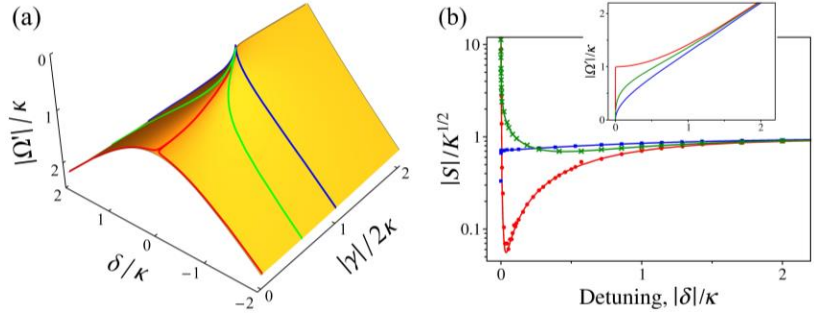

FIG. 5. (a) Beat frequency $|\Omega'|$ and (b) $|S|/K^{1/2}$ vs. detuning for conservative coupling ($\kappa'' = 0$) at varying levels of saturation imbalance: $\psi_0 = 0$ (blue), $\psi_0 = |\partial_\delta \gamma_{1,0}|$ (green), $\psi_0 = 2|\partial_\delta \gamma_{1,0}|$ (red). The gain in the first resonator $-\gamma_1$ is set to saturate, and the saturation imbalances are achieved by setting the second resonator so that it experiences an equal amount of saturable gain (blue), constant gain (green), and an equal amount of saturable loss (red). In each case the system saturates to threshold, and the nonsaturable losses are set so the threshold corresponds to the EP at $\delta = 0$. The data points are determined from numerical solutions of the nonlinear coupled-mode equations, while the curves are analytic solutions of Eq. (18) (the red curve requires determination of $\psi_0$ from the numerical computation). The inset in (b) shows the projections of the curves in (a). The unsaturated parameters are: $\gamma_1^u/\kappa' = \gamma_2^u/\kappa' = -1.8/1.4$ (blue); $\gamma_1^u/\kappa' = -2.5$ and $\gamma_2^u = 0$ (green); and $\gamma_1^u/\kappa' = -\gamma_2^u/\kappa' = -1.5$ (red), corresponding to the saturation imbalances above.

In contrast, for maximally dissipative coupling $\kappa' = 0$. Moreover, the saturation is always balanced ($\psi_0 = 0$) because $\gamma_0(\delta) = 0$ for all values of $\delta$ regardless of how far the system is pumped above threshold. Therefore, the condition for enhancement, Eq. (21), cannot be satisfied in this case. Indeed, outside the deadband $\Omega' = [\delta^2 - (\kappa'')^2]^{1/2}$ and $S = \delta/\Omega'$ (from Eq. (7) with interchanged parameters). In addition, since $B_0 = \gamma_0 \delta = 0$, the threshold condition is $A_0 = \delta^2 - (\kappa'')^2 = -\gamma_{avg,0}^2$. Substituting $|\tilde{\Omega}|^2 = (A_0^2 + B_0^2)^{1/2} = A_0$ into Eq. (13) we have $K = (\delta/\Omega')^2$ which agrees with Eq. (10) with interchanged parameters. In other words, $S$ and $K$ are independent of $\gamma$, so gain saturation does not affect them and there is no variation from the linear case.

Therefore, gain saturation can remove the rotational invariance of EPs, and studies performed near the EP at the deadband edge should not be expected to be indicative of the precision at the PT-symmetric EP. In fact, gain saturation can break the rotation symmetry even when there is no saturation imbalance. Near the maximally dissipative EP (the deadband edge) gain saturation distorts the beat signal resulting in harmonics that pull the beat frequency to larger values than that expected from Eq. (6) [37]. The same distortion is not expected to occur near the conservative coupling (PT-symmetric) EP [46].

## VI. ABOVE THRESHOLD BEHAVIOR: LINEARIZATION APPROACH

The results above still rely on the use of the Petermann factor, which is fundamentally linear. We've simply substituted saturated gain coefficients for nonsaturable ones in Eq. (8). Consequently, we cannot discount the possibility that a proper treatment of the excess noise might eliminate the enhancement illustrated in Fig. 5. The nonlinear coupled-mode equations can be made stochastic by the addition of Langevin noise terms to Eq. (1). These equations have already been solved for the case of maximally dissipative coupling [42, 52, 57]. Outside the deadband ($|\delta|>|\kappa''|$) where two modes lase and beat together, the laser phase obeys the Adler equation. This equation is anharmonic owing to the gain saturation, resulting in the appearance of harmonics in the beat-note spectrum as mentioned above [35, 37, 42, 57]. Injecting quantum noise into the Adler equation via a Langevin term and calculating the fundamental linewidth via the Fokker-Planck equation results in a small correction factor, $1+(\kappa'')^2/2\delta^2$, which should be multiplied by the Petermann factor obtained from Eq. (10), $\delta^2/[\delta^2-(\kappa'')^2]$ [42, 51, 57]. Note that this factor is never larger than $3/2$, taking its largest value at the EP. Indeed, experiments performed in this region have shown that the linear theory works remarkably well [42, 52].

For frequencies inside the deadband $|\delta|<|\kappa''|$, on the other hand, only one mode lases and the beat frequency is zero, i.e., there is a stable steady state. In this region the nonlinear coupled-mode equations, which include the Langevin noise terms, can be linearized around this steady-state. The excess noise can then be determined from the noise power spectrum [58]. Using this approach, Van der Lee et al. have shown that while the intensity excess-noise follows the linear Petermann factor (with nonsaturable coefficients), the phase noise is strongly modified by the saturation, and can be greatly reduced at high steady-state laser intensities [58]. This result demonstrates that when the system is pumped above threshold, the excess-noise factor can deviate significantly from the linear prediction, even with the use of saturated coefficients. Unfortunately, the reduction in excess noise is not beneficial to the application discussed here, because the scale factor and thus the precision are zero within the deadband region.

Nevertheless, we can take a similar approach to derive the excess-noise factor for conservative coupling in the single mode, a.k.a., lasing-without-gain (LWG) [46], regime. The approach is valid in the good cavity limit, where the polarization and inversion can be adiabatically eliminated. We assume the addition of the lossy resonator does not invalidate this approximation. In addition, the system is assumed to be not too far above threshold so that noise in the inversion can be neglected, leaving spontaneous emission as the dominant quantum noise source. Under these conditions, a single Langevin noise term can be used in Eq. (3). The presence of the loss coefficients in Eq. (3) makes the procedure a bit more complicated than for the maximally dissipative coupling case (which assumes $\gamma_1=\gamma_2=0$), but it proceeds along similar lines. We first make the approximation $\gamma_i(I_i)=\gamma_i^s(I_i)+\gamma_i^n \approx \hat{\gamma}_i^u - \hat{\beta}_i I_i$ $(i=1,2)$, where $\hat{\beta}_i = \beta_i \gamma_i^u$ and $\hat{\gamma}_i^u = \gamma_i^u + \gamma_i^n$. We then transform to a more convenient basis consisting of the orthogonal variables $2\theta=\phi_1-\phi_2$, $2\phi=\phi_1+\phi_2$, $\tan\chi=(\mathcal{E}_1-\mathcal{E}_2)/(\mathcal{E}_1+\mathcal{E}_2)$, and $I=\mathcal{E}_1^2+\mathcal{E}_2^2$, by setting $e_i = \mathcal{E}_i \exp(i\phi_i)$ as shown in [58], to obtain the following coupled equations

$$\dot{\theta} = (\delta/2) - (\kappa'/2)\tan 2\chi \cos 2\theta + f_\theta, \tag{23a}$$

$$\dot{\phi} = (\kappa'/2)\cos 2\theta/\cos 2\chi + f_\phi, \tag{23b}$$

$$\dot{\chi} = (\kappa'/2)\sin 2\theta - (\gamma/4)\cos 2\chi + f_\chi, \tag{23c}$$

$$\dot{I}/2I = -(\gamma_{avg}/2) - (\gamma/4)\sin 2\chi + f_I, \tag{23d}$$

along with the additional condition

$$\gamma_{avg} = -\kappa' \tan 2\chi \sin 2\theta - (\dot{I}/I - 2\dot{\chi}\tan 2\chi), \tag{24}$$

where $\gamma = \hat{\gamma}^u - \beta_d I - \beta_s I \sin 2\chi$, $\gamma_{avg} = \hat{\gamma}_{avg}^u - (\beta_d I/2)\sin 2\chi - \beta_s I/2$, $\hat{\gamma}^u = \hat{\gamma}_1^u - \hat{\gamma}_2^u$, $\hat{\gamma}_{avg}^u = (\hat{\gamma}_1^u + \hat{\gamma}_2^u)/2$, $\beta_d = (\hat{\beta}_1 - \hat{\beta}_2)/2$, and $\beta_s = (\hat{\beta}_1 + \hat{\beta}_2)/2$. The Langevin noise terms

$\{f_\theta, f_\phi, f_\chi, f_I\}$ are now characterized by the correlation relations $\langle f_i(t)^* f_j(t')\rangle = \mathcal{D}_i \delta_{ij}\delta(t-t')$, where

$$\mathcal{D}_I = (I_1 R_1^{sp} + I_2 R_2^{sp})/2I^2, \tag{25a}$$

$$\mathcal{D}_\chi = (I_2 R_1^{sp} + I_1 R_2^{sp})/2I^2, \tag{25b}$$

$$\mathcal{D}_\theta = \mathcal{D}_\phi = \frac{R_1^{sp}/I_1 + R_2^{sp}/I_2}{8\sin^2 2\theta}. \tag{25c}$$

Note that these autocorrelations differ from those for the conventional gyroscope [58] because the resonators can be physically distinct with different intensities and spontaneous emission rates.

If, for now, we set only the time derivatives for the intensities equal to zero, i.e., $\dot\chi = \dot I = \dot I_1 = \dot I_2 = 0$, we obtain the steady-state solutions $\{I_{1,0}, I_{2,0}, \chi_0\}$

$$\gamma_{1,0} I_{1,0} = -\gamma_{2,0} I_{2,0}, \tag{26a}$$

$$\gamma_0/2 = \kappa' \sin 2\theta / \cos 2\chi_0, \tag{26b}$$

$$\gamma_{avg,0} = -\kappa' \tan 2\chi_0 \sin 2\theta. \tag{26c}$$

From the latter two equations we obtain $(\kappa' \sin 2\theta)^2 = (\gamma_0/2)^2 - (\gamma_{avg,0})^2 = -\gamma_{1,0}\gamma_{2,0}$, which is equivalent to the condition for threshold at an arbitrary detuning $\delta$. In other words, the system saturates to the threshold loss values $\{\gamma_{1,0}, \gamma_{2,0}, \gamma_0, \gamma_{avg,0}\}$ in the steady state. These steady-state relations apply when $|\kappa' \sin 2\theta| < |\gamma_0/2|$. By equating the threshold condition above with Eq. (22) we find that $\sin^2(2\theta) = [1+(\delta/\gamma_{avg,0})^2]^{-1}$. It then follows that $\kappa' \tan 2\chi_0 \cos 2\theta = \delta$ such that $\dot\theta = 0$. Therefore, in this regime whenever the intensity is in steady state the phase difference must also be in steady state. It is not required that the overall phase $\phi$ be in steady state to obtain this result. In fact, $\dot\phi = 0$ only under the additional conditions

$$\begin{aligned}\cos 2\theta_0 &= 0\\ \delta &= 0\end{aligned} \tag{27}$$

Thus, while $\chi$, $I$, and $\theta$ can be in steady-state at nonzero detunings, $\phi$ is only stationary when $\delta = 0$ (similarly, for dissipative coupling the phase is stationary only when $\gamma = 0$). These fully steady-state conditions apply when $|\kappa'| < |\gamma_0/2|$, i.e., for subexceptional couplings. Under these conditions $(\gamma_{avg,0})^2 = (\kappa' \tan 2\chi_0)^2 = (\gamma_0/2)^2 - (\kappa')^2$, which corresponds to the threshold condition at $\delta = 0$. Therefore, unlike the other variables, we will not be able to find an analytic form for the excess noise for the overall phase except at $\delta = 0$.

Linearizing around the stable steady-state, i.e., making the substitutions $\theta = \theta_0 + \Delta\theta$, $\chi = \chi_0 + \Delta\chi$, and $I = I_0(1+2\Delta I)$ results in the following coupled equations

$$\Delta\dot\theta = \gamma_{avg,0} \Delta\theta + p\Delta\chi + f_\theta \tag{28a}$$

$$\dot\phi = \gamma_0 \Delta\theta/2 + q\Delta\chi + f_\phi \tag{28b}$$

$$\Delta\dot\chi = A\Delta\chi + B\Delta I + f_\chi \tag{28c}$$

$$\Delta\dot I = C\Delta\chi + D\Delta I + f_I \tag{28d}$$

where $p = -2\delta/(\sin 2\chi_0 \cos 2\chi_0)$, $q = \delta\gamma_0/(\kappa' \sin 2\theta_0)$, $A = (\hat\gamma^u - \gamma_0)\gamma_{avg,0}/\gamma_0 + (\hat\gamma^u_{avg} - 2\gamma_{avg,0})$, $B = \kappa' \sin 2\theta_0 (\hat\gamma^u - \gamma_0)/\gamma_0$,

$C = \kappa' \sin 2\theta_0 (\hat{\gamma}^u - 2\gamma_0)/\gamma_0$, and $D = -(\hat{\gamma}^u - \gamma_0)\gamma_{avg,0}/\gamma_0 + (\hat{\gamma}^u_{avg} - \gamma_{avg,0})$. Thus, for the amplitude variables $\{\chi, I\}$ the dependency on detuning occurs implicitly via $\gamma_0$ and $\gamma_{avg,0}$, and explicitly through $\sin 2\theta_0$, whereas for the phase variables $\{\theta, \phi\}$ there is an additional explicit dependency owing to the $p$ and $q$ terms. Note also that the factors $A$, $B$, $C$, and $D$ are implicitly dependent on intensity via the unsaturated quantities $\hat{\gamma}^u$ and $\hat{\gamma}^u_{avg}$, indicating that the excess noise will depend on how far above threshold the system is pumped.

Taking the Fourier transform and applying the Wiener-Khintchine theorem yields the noise power spectra

$$\langle |\Delta\theta(\omega)|^2 \rangle = \frac{1}{\omega^2 + \gamma_{avg,0}^2}\left[\mathcal{D}_\theta + p^2 \langle |\Delta\chi(\omega)|^2 \rangle\right] \tag{29a}$$

$$\langle |\phi(\omega)|^2 \rangle = \frac{1}{\omega^2}\left[\frac{(\gamma_0/2)^2}{\omega^2 + \gamma_{avg,0}^2}\left(\mathcal{D}_\theta + p^2 \langle |\Delta\chi(\omega)|^2 \rangle\right) + \mathcal{D}_\phi + q^2 \langle |\Delta\chi(\omega)|^2 \rangle\right] \tag{29b}$$

$$\langle |\Delta\chi(\omega)|^2 \rangle = \frac{(D^2 + \omega^2)\mathcal{D}_\chi + B^2 \mathcal{D}_I}{(AD - BC - \omega^2)^2 + (A + D)^2 \omega^2} \tag{29c}$$

$$\langle |\Delta I(\omega)|^2 \rangle = \frac{(A^2 + \omega^2)\mathcal{D}_I + C^2 \mathcal{D}_\chi}{(AD - BC - \omega^2)^2 + (A + D)^2 \omega^2}. \tag{29d}$$

Note that: (i) due to the mutual coupling between $\Delta\chi$ and $\Delta I$, $f_I$ projects onto $\Delta\chi$ and $f_\chi$ projects onto $\Delta I$, (ii) these noise sources only project onto the phase variables when $\delta \neq 0$, and (iii) there is a one way coupling between $\Delta\theta$ and $\phi$, so $f_\theta$ projects onto $\phi$ but $f_\phi$ does not project onto $\Delta\theta$. As we will see, the one-way coupling between the phase variables leads to only a small departure from the quasilinear Petermann result, whereas the mutual coupling of the amplitude variables leads to a much larger deviation.

The excess-noise factors are then found by taking the ratio of the noise spectrum at an arbitrary coupling to that at zero coupling ($\kappa' \to 0$). As we are primarily interested in the behavior near the EP, in what follows we will set $\delta = 0$ such that all the variables reach steady state and the terms containing $p$ and $q$ vanish. For the phase difference this results in

$$K_\theta(\omega) = \eta_\theta \left[\frac{\omega^2 + (\gamma_0/2)^2_{\kappa'=0}}{\omega^2 + \gamma_{avg,0}^2}\right] = \eta_\theta K_\omega, \tag{30}$$

where

$$\eta_\theta = \frac{\mathcal{D}_\theta}{\mathcal{D}_\theta|_{\kappa'=0}} \tag{31}$$

is an intensity dependent factor arising from the modified autocorrelation in the coupled system. At $\omega = 0$ Eq. (30) reduces to $K_\theta(0) = (\eta_\theta \Gamma)K$, where $\Gamma = (\gamma_0/2)^2_{\kappa'=0}/(\gamma_0/2)^2$ and $K = (\gamma_0/2\gamma_{avg,0})^2$ is the usual Petermann factor (with saturated coefficients). Thus, the factor $\eta_\theta \Gamma$ modifies the excess noise factor from that predicted by the quasilinear result. Assuming spontaneous emission in only the first resonator ($R_2^{sp} = 0$), we obtain $\eta_\theta = [I_1]_{\kappa'=0}/I_1$, which asymptotes to its minimum value of $\eta_\theta = 2$ at the EP in the limit of large pumping. If, in addition, equal nonsaturable losses are assumed ($\gamma_1^n = \gamma_2^n$) we find that $\Gamma = 1/4$ at the EP, regardless of the level of pumping. These deviations from the Petermann result occur because at a particular pumping rate, characterized by $R_1^{sp}$, the steady-state coefficients ($\Gamma$) and intensities ($\eta_\theta$) must be different for the coupled and uncoupled systems. Note that $\Gamma K$ is exactly what we would expect to find had we properly generalized the Petermann factor (see top of Eq. (10)), to account for the different saturation behaviors, while the additional factor $\eta_\theta$ arises from the fact that the conventional linewidths of the two systems are different.

For the overall phase, at $\omega = 0$, we obtain

$$K_\phi(0) = \eta_\theta \left[ \frac{(\gamma_0/2)^2 - (\kappa')^2/2}{(\gamma_0/2)^2 - (\kappa')^2} \right]. \tag{32}$$

The factor in brackets differs from the Petermann result by the factor $1 - (2\kappa'/\gamma_0)^2/2$, which has a maximum of 1/2 at the EP (where $|\kappa'| = |\gamma_0/2|$), i.e., $K_\phi^{EP}(0) = \eta_\theta K/2$. Thus, for the phase variables $\{\theta, \phi\}$ the dependence of the excess noise on laser intensity occurs solely through the factor $\eta_\theta$. As we have already pointed out, the same is not true for the intensity quantities $\{\chi, I\}$ whose intensity dependence also arises from the factors $A$, $B$, $C$, and $D$.

For the amplitude variables, setting $\kappa' = 0$ in Eqs. (28c) and (28d), uncouples the equations ($B = C = 0$) resulting in the noise spectrum for the higher-gain (first) resonator

$$\langle |\Delta I(\omega)|^2 \rangle = \left[ \frac{\mathcal{D}_I}{\omega^2 + D^2} \right]_{\kappa'=0}, \tag{33}$$

where $D|_{\kappa'=0} = -\gamma_1^n \hat{\gamma}_1^u / \gamma_1^u$. Unlike Eq. (29d), in this case Eq. (33) is exact and does not rely on the low-intensity approximation ($\beta_i I_i \ll 1$). Therefore, the intensity excess noise factor at $\omega = 0$ is

$$K_I(0) = \frac{A^2 \mathcal{D}_I + C^2 \mathcal{D}_\chi}{(AD - BC)^2} \left[ \frac{D^2}{\mathcal{D}_I} \right]_{\kappa'=0}. \tag{34}$$

Note that because of the coupling, the divergence is now determined by the zero in the determinant rather than by being at the EP. At the EP, Eq. (34) reduces to

$$K_I^{EP} = \eta_I \left[ \frac{(2\hat{\gamma}_{avg}^u)^2 + (\hat{\gamma}^u - 2\gamma_0)^2}{(2\hat{\gamma}_{avg}^u)^2 - (\hat{\gamma}^u - \gamma_0)(\hat{\gamma}^u - 2\gamma_0)} \right] \left( \frac{\gamma_1^n \hat{\gamma}_1^u}{\gamma_1^u} \right)^2, \tag{35}$$

where $\eta_I = \mathcal{D}_I / [\mathcal{D}_I]_{\kappa'=0}$. Importantly, $K_I^{EP}$ does not necessarily diverge but remains finite for a wide set of $\{\gamma_1^u, \gamma_2^u\}$ values. An exception is when the system is pumped at (not above) threshold. In this case $K_I^{EP}$ diverges because the term in square brackets on the RHS reduces to the Petermann result, and in addition $\eta_I$ diverges. In the limit of large pumping, on the other hand, $\eta_I = 1/2$ at the EP (assuming $R_2^{sp} = 0$).

To verify the results of the linearization procedure described above, we numerically solved the coupled nonlinear equations as described in Fig. 5, but now with noise injected via Langevin terms added to the real and imaginary parts of Eq. (1) to obtain $K_\theta$ and $K_I$. This involves computing $\Delta\theta = \theta - \theta_0$ and $\Delta I = (I/I_0 - 1)/2$, taking the fast Fourier transform to obtain Eqs. (29a) and (29d), then repeating the procedure and normalizing to the results obtained at $\kappa' = 0$. This was repeated at a variety of different couplings and detunings. We assume in these calculations that only the first resonator saturates and that there is no spontaneous emission in the second resonator, i.e., $R_2^{sp} = 0$. Thus, $\eta_\theta = [I_1]_{\kappa'=0}/I_1$ and $\eta_I = [I]_{\kappa'=0}/2I$. In addition $K_\theta$ and $K_I$ were calculated from the same data, i.e., the same noise sources and value of $R_1^{sp}$ were used. The system was pumped only slightly above threshold ($\hat{\gamma}^u/\gamma_0 = 1.02$) to ensure the validity of the low-intensity approximation at the EP. In addition, as in the green curve of Fig. 5, we set $\gamma_1^n = \gamma_2^n = \kappa'$ and $\gamma_2^u = 0$, which allows convenient analytic determination of the intensity $I_0$.

In Fig. 6, $K_\theta$ is plotted against frequency, coupling, and detuning. The detuning is constant ($\delta = 0$) in Figs. 6(a)-(c), whereas the coupling is held constant ($\kappa' = 0.999\kappa^{EP}$) in Fig. 6(d). The linearized model (solid curves) matches the numerical computation in each of the individual figures. In Fig. 6(a) noise power spectra are plotted for three different coupling values, from which the curves of $K_\theta$ vs. frequency are constructed in Fig. 6(b). Each spectrum represents an average of 30 individual spectra. Note that the phase noise in Fig. 6(a) is colored, being dominated by white noise at low frequencies and random walk at high frequencies. The transition between these regimes occurs at $\omega = |\gamma_{avg,0}|$, which redshifts as the coupling increases, approaching $\omega = 0$ at the EP. Consequently, for the top curve ($\kappa' = 0.999\kappa^{EP}$) the transition frequency is too low to observe, as it falls below the smallest frequency, $\omega_\alpha$, in the spectrum, determined by the duration of our numerical simulation. Note also that in Fig. 6(b) $K_\theta$ diverges in the limit $\omega \to 0$ as the EP is approached but otherwise remains finite. In Figs. 6(c) and 6(d) $K_\theta$ is plotted vs. coupling and detuning, respectively, at $\omega_\alpha$. The dashed indigo curve in Fig. 6(c) is a plot of $K_\omega$, whereas the green dotted curve is the quasilinear theory, $K$. The data exceeds $K_\omega$ by the factor $\eta_\theta$, and also exceeds $K$ even when $\kappa' = 0.999\kappa^{EP}$, because the system is far below the strong pumping limit. But even in this weak pumping case, for the particular case of $\kappa' = \kappa^{EP}$, $K$ diverges and thus exceeds $K_\theta$ (which, again, remains finite even at the EP because of the nonzero frequency $\omega_\alpha$). In Fig. 6(d) the linearized model matches the data at small $\delta$, but at higher detunings the threshold drops and the intensity grows sufficiently high that the low-intensity approximation is no longer valid. Again, the green dotted curve shows the quasilinear theory underestimates the data owing to the weak pumping and subexceptional coupling ($\kappa' = 0.999\kappa^{EP}$).

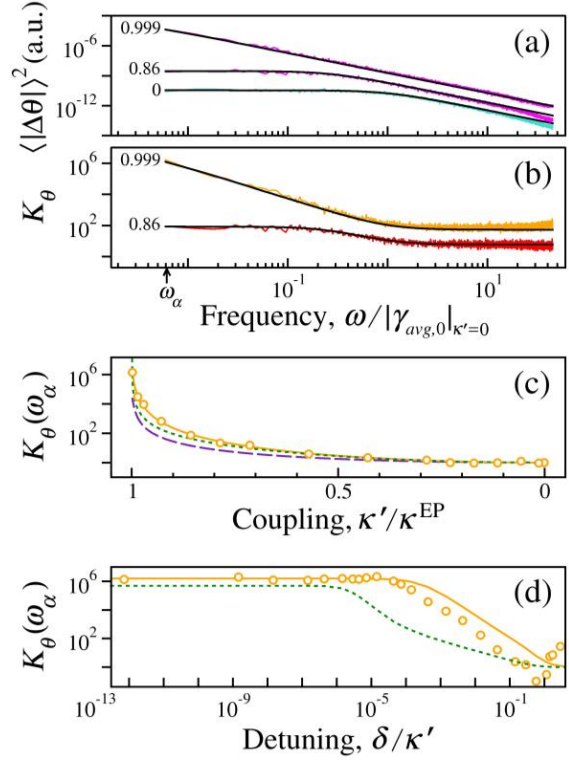

FIG. 6. Phase excess-noise factor, $K_\theta$, vs. frequency (b), coupling (c), and detuning (d) as obtained from numerical solution of the nonlinear coupled-mode equations including Langevin noise terms, and compared with the analytic solution obtained from linearizing around the steady-state (solid curves). In (a)-(c) the detuning is $\delta = 0$, while in (d) the coupling is $\kappa' = 0.999\kappa^{EP}$. In (a) noise power spectra are shown for three values of $\kappa'/\kappa^{EP}$ {0, 0.86, 0.999}, from which the curves in (b) are constructed. In (c) and (d) $K_\theta$ is plotted against $\kappa'/\kappa^{EP}$ and $\delta$, respectively, at $\omega_\alpha$, the smallest frequency present in the spectrum in (b). The dashed indigo curve is $K_\omega$, whereas the dotted green curve is the Petermann prediction, $K$. The frequencies along the horizontal axes in (a) and (b) are normalized to $|\gamma_{avg,0}|_{\kappa'=0}$ such that the white-noise to random-walk transition for the bottommost ($\kappa'=0$) curve in (a) occurs at unity.

Fig. 7 shows the same sort of plots for $K_I$. The parameters are the same as in Fig. 6, except that in Fig. 7(c) an additional set of data (blue curve and crosses) is included to demonstrate the effect of pumping farther above threshold ($\hat{\gamma}^u/\gamma_0 = 1.25$, same pumping level as green curve in Fig. 5). One key difference from the data in Fig. 6 is that $K_I$ does not diverge at the EP. Thus, in Figs. 6(c) and 6(d) the zero-frequency value of $K_I$ can be plotted. Moreover, in Fig. 7(c) the linearized model only matches the data near the EP. This is because the intensity $I_0$ grows as the coupling decreases to the point that the low-intensity approximation is no longer valid. This is underscored by the results at stronger pumping where the intensity is higher and the agreement is even worse. Note that in both cases $K_I$ remains finite even at the EP and is lower at stronger pumping in accordance with Eq. (35). In addition, the coloring of the intensity excess noise is much different than in Fig. 6. A resonance peak appears in the spectrum owing to the zero in the denominator of Eqs. 29(c-d). The peak becomes stronger, moves to higher frequencies, and broadens as the pumping increases, characteristic of a relaxation oscillation.

In this case, the oscillation is induced by the quantum noise, and would not be observable if not for the enhancement near the EP as shown in Fig. 6(a).

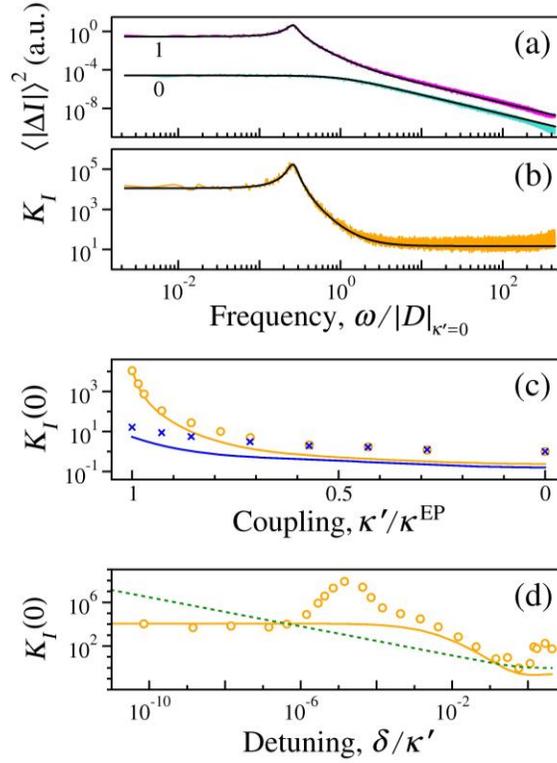

FIG. 7. Noise power spectra (a) and intensity excess-noise factor, $K_I$, vs. frequency (b), coupling (c), and detuning (d), compared with the linearized theory (solid curves). In (a) the value of $\kappa'/\kappa^{EP}$ is indicated under the curves (1 and 0). The curve in (b) is the ratio of the two curves in (a). In (b) the system saturates to the EP, corresponding to the leftmost data points in (c) and (d). In (c) and (d) only the zero-frequency value of $K_I$ is plotted. In (c) the effect of pumping farther above threshold is shown (blue curve and crosses), which decreases the excess noise factor. The intensity is sufficiently high in this case that the analytic solution no longer predicts the data, even at the EP. The dashed curve in (d) is the quasilinear prediction, i.e., the Petermann factor $K$ with saturated coefficients, which diverges at the EP. Note that $K_I$ remains finite and there is a region of detunings near the EP where $K_I \leq K$. In this case the frequencies in (a) and (b) are normalized to $|D|_{\kappa'=0}$ in accordance with Eq. (33).

In Fig. 7(d) the effect of detuning is shown. A large peak occurs that is not predicted by the linearized model. This is again due to the breakdown of the low-intensity approximation as the detuning increases. In particular, the mutual coupling in $K_I$ requires lower intensities to satisfy the approximation than is the case for $K_\theta$. The quasilinear model is also shown (dashed curve). The two models cross, with the linear model overestimating the noise at small detunings and underestimating it at larger detunings. Increasing the unsaturated gain $-\gamma_1^u$ results in higher intensities, $I_0$, which reduces the value of $K_I^{EP}$ (at $\delta = 0$) and moves the crossing point to higher detunings. Hence, there is a region near the EP where the intensity excess noise is lower than expected from the quasilinear theory ($K_I \leq K$) over a bandwidth and by an amount that increases with increased pumping.

In summary, in conservatively coupled systems both phase and intensity excess noise factors can be decreased through increased pumping via the factor $\eta$ and by the coupling of the noise fluctuations, which has a greater effect on $K_I$. At the EP $K_I$ remains finite even at $\omega = 0$, decreasing with higher steady-state intensities, while $K_\phi$ and $K_\theta$ still diverge at $\omega = 0$ but are multiplied by factors of $\eta_\theta / 2$ and $\eta_\theta \Gamma$, respectively. In the strong pumping limit $\eta_\theta = 2$ such that $K_\phi = K$ and $K_\theta = 2K_\omega = K/2$ (assuming equal nonsaturable losses). This reduction of the excess noise with pumping in shown in Fig. 8, where $K_\theta$ and $K_I$ are plotted near the EP vs. the pumping level relative to threshold. For small pumping levels the excess noise is higher than the quasilinear prediction (which does not change with pumping) because $\eta > 1$, but as the pumping increases $K_I$ falls below unity whereas $K_\theta$ approaches $2K_\omega$. Note that $K_\omega$ falls farther below the quasilinear result than the factor $\Gamma$ would predict, i.e., $K_\omega < \Gamma K$, due to the nonzero frequency $\omega_\alpha$. In both cases the data is better represented by the linearized nonlinear model and is reduced below the quasilinear prediction. Our use of saturated coefficients in Eqs. (18)-(22), therefore, appears to be a worst case scenario. At the small detunings of interest for gyroscopes, the precision can be even higher than predicted by the quasilinear analysis.

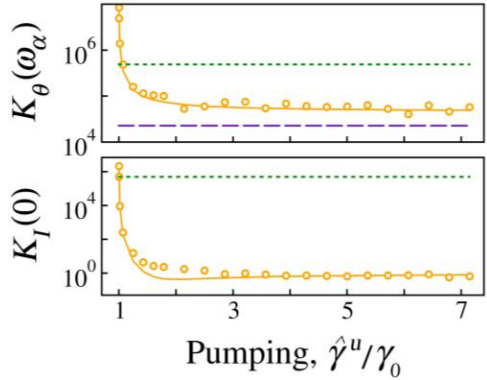

FIG. 8. Excess noise factors near the EP ($\delta = 0$, $\kappa' = 0.999\,\kappa^{EP}$) vs. relative pumping level. Numerical data (open circles) is compared with the linearized (solid orange curve) and quasilinear (green dotted) models, as well as with $K_\omega$ (dashed indigo). The nonsaturable gain was equal for the two resonators and the pumping level was varied by increasing the saturable gain in the first resonator, with no saturable gain included in the second resonator.

These results for conservative coupling are complementary to the results found previously for maximally dissipative coupling that the phase excess noise can be reduced to unity in the limit of strong pumping, while the intensity noise is unchanged from the Petermann result [58]. In contrast, here the intensity noise is significantly reduced while the phase noise is less affected. In both cases, the deviation from the Petermann result at strong pumping occurs because fluctuations in the orthogonal variables are coupled. In the case of dissipative coupling the phase $\{\phi, \Delta\theta\}$ and amplitude $\{\Delta\chi, \Delta I\}$ variables are coupled, i.e., $\Delta\chi$ is coupled to $\phi$ and $\Delta\theta$ is coupled to $\Delta I$. The reduced phase excess noise comes about as a result of the coupling with the amplitude fluctuations, which are suppressed due to gain saturation, especially far above threshold. The same thing happens in conservatively coupled systems, except in this case there is greatly reduced amplitude to phase coupling at ($\Delta\theta$ is coupled to $\phi$ and $\Delta\chi$ is coupled to $\Delta I$ at $\delta = 0$, see Eq. 28), so the noise suppression is isolated to the intensity excess noise. Essentially, the saturation acts as a negative feedback on the amplitude fluctuations and this can also affect the phase excess noise in situations where there is amplitude to phase coupling. The decoupling of the phase and amplitude noise components in conservatively coupled systems also leads to a sinusoidal temporal response (instead of the anharmonic response that characterizes dissipative coupling) and elimination of harmonics from the beat spectrum (subsequent to its recovery as described below). Furthermore, whereas the one-way coupling of $\phi$ to $\Delta\theta$ leads to only a small change from the Petermann result, the mutual coupling of the amplitude variables leads to a dramatic reduction in the intensity excess noise. This mutual coupling does not occur for dissipative coupling [58] and explains why the intensity excess noise factor can fall well below unity under strong pumping.

## VII. MEASUREMENT PRECISION FOR COLORED QUANTUM NOISE

The metric $S/K^{1/2}$, established in section III for the measurement precision, relies on the assumption that the frequency noise in both the uncoupled and coupled systems is white. However, we observed in the previous section that the excess quantum noise is colored. The coloring of the noise, as shown below in Fig. 9, thus complicates the determination of the precision. For laser gyroscopes the relevant excess noise factor is for the relative phase, i.e., $K_\theta$. Note that when $p = 0 = \delta$, Eq. (29a) describes Gauss-Markov noise with a characteristic decay rate given by $|\gamma_{avg,0}|$. There are three limiting regions identified in the figure. In region I the phase noise in both the coupled and uncoupled systems is white because $\omega^2 \ll \gamma^2_{avg,0}$ (which also ensures that $\omega^2 \ll \gamma^2_{avg,0}\big|_{\kappa'=0}$). In this limit the single-sided noise power spectral density (PSD), determined from Eq. (29a), reduces to $S_\theta = \mathcal{D}_\theta / \gamma^2_{avg,0}$. Converting to a frequency noise PSD we have $S_\omega = \omega^2 S_\theta = \mathcal{D}_\theta (\omega/\gamma_{avg,0})^2$, corresponding to an Allan variance $\sigma^2_\omega = 3 f_h S_\omega$, where $f_h$ is the cutoff frequency of the low-pass filter (the highest possible measurement rate). Assuming $f_h$ is the same for the two systems, the relative error for a frequency measurement is then given by $\varepsilon \equiv \sigma_\omega / [\sigma_\omega]_{\kappa'=0} = K_\theta^{1/2} = (\eta_\theta K_\omega)^{1/2}$ and we recover our original metric for the enhancement in precision $S/\varepsilon = S/K_\theta^{1/2}$.

In region III we are in the opposite limit where $\omega^2 \gg \gamma^2_{avg,0}\big|_{\kappa'=0}$ (which also ensures that $\omega^2 \gg \gamma^2_{avg,0}$), and the phase noise is dominated by random walk (equivalent to white frequency noise) in both systems. In this case $S_\theta = \mathcal{D}_\theta / \omega^2$ and $S_\omega = \omega^2 S_\theta = \mathcal{D}_\theta$, corresponding to $\sigma^2_\omega = S_\omega / 2\tau$, where $\tau = 1/\omega$ is the measurement time. For white frequency noise the linewidth can be obtained analytically to be $\pi S_\omega$. Thus, in this region we can explicitly associate the increased error with an increase in the laser linewidth. The measurement times cancel leaving $\varepsilon = \eta_\theta^{1/2} = K_\theta^{1/2}$ since in this limit $K_\omega$ is unity. In region II, on the other hand, random walk phase noise dominates in the coupled system ($\omega^2 \gg \gamma^2_{avg,0}$), but white phase noise dominates in the uncoupled system ($\omega^2 \gg \gamma^2_{avg,0}\big|_{\kappa'=0}$). In this limit the times and cutoff frequencies do not cancel so $\varepsilon$ is no longer simply the ratio of the square root of the PSDs for the two systems. Instead, we have $\varepsilon = (\mathcal{D}_\theta / 2\tau)^{1/2} / (3 f_h \mathcal{D}_\theta \omega^2 / \gamma^2_{avg,0})^{1/2}_{\kappa'=0}$, i.e., the relative error becomes frequency dependent. In summary, we have:

$$\varepsilon = \begin{cases} K_\theta^{1/2} = (\eta_\theta K_\omega)^{1/2} & \omega^2 \ll \gamma^2_{avg,0} \\ \eta_\theta^{1/2} \left|\gamma_{avg,0}\right|_{\kappa'=0} (\pi / 3\omega\omega_h)^{1/2} & \gamma^2_{avg,0} \ll \omega^2 \ll \gamma^2_{avg,0}\big|_{\kappa'=0} \\ K_\theta^{1/2} = \eta_\theta^{1/2} & \omega^2 \gg \gamma^2_{avg,0}\big|_{\kappa'=0} \end{cases}, \quad (36)$$

where $\omega_h = 2\pi f_h$.

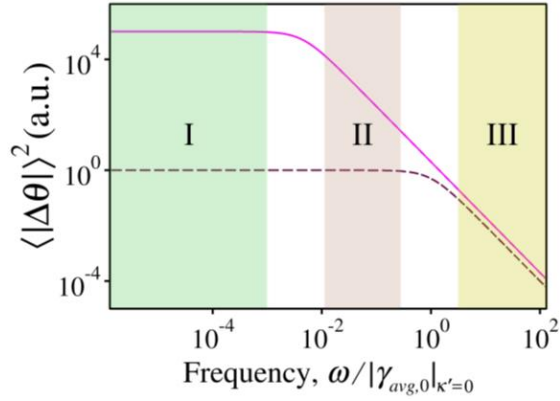

FIG. 9. Phase noise power spectrum for an uncoupled ($\kappa' = 0$, dashed brown curve) and a coupled system near the EP ($\kappa'/\kappa^{EP} = 1$, $\delta/\kappa' = 1.5\times 10^{-4}$, solid magenta curve) in the strong pumping limit ($\hat{\gamma}^u \gg \gamma_0$). Region I: white phase noise dominates in both systems. Region II: random walk phase noise (white frequency noise) dominates in the coupled system, but white phase noise dominates in the uncoupled system. Region III: random walk phase noise dominates in both systems. The unshaded areas are transition regions. The vertical axes are normalized such that the power spectrum of the uncoupled system is unity in region I.

Note, however, that although $\varepsilon$ diverges at the EP ($\delta = 0$, $\kappa' = \kappa^{EP}$), the decay rate simultaneously drops to $\gamma_{avg,0} = 0$, such that region I becomes infinitely small at $\omega = 0$. Thus, near an EP, the predominant mode of operation at low frequencies will be in region II because measurement frequency cannot be zero. In addition, according to the middle equation above, the error decreases as the frequency increases, resulting in larger enhancements in precision at higher values of $\omega$. If we set the measurement rate to the maximum, i.e., $\omega = \omega_h$, we find in region II that

$$\varepsilon = (\eta_\theta K_\omega)^{1/2}(\pi/3)^{1/2} \approx K_\theta^{1/2}, \qquad (37)$$

where we've used the fact that $K_\omega = (\gamma_0/2)^2_{\kappa'=0}/\omega^2$ in region II. Thus, we find that our original metric $S/K_\theta^{1/2}$ provides a good estimate for the enhancement in precision for all three regions.

In Fig. 10 we plot the precision $\varsigma \equiv S/\varepsilon$ as obtained from the top (dashed orange curve) and middle (solid blue curve) equations of Eq. (36) versus the detuning. The region of validity for the dashed orange curve is shaded green (region I), while that for the solid blue curve is shaded red (region II). The top two figures are slightly subexceptional ($\kappa'/\kappa^{EP} = 0.9999$), while the bottom figures are at $\kappa'/\kappa^{EP} = 1$. In addition, higher(lower) frequencies are shown on the right (left) to demonstrate the effect of increasing the measurement rate. The frequencies are set to $\omega/|\gamma_{avg,0}|_{\kappa'=0} = 1.32\times 10^{-5}/0.7$ (left) and $\omega/|\gamma_{avg,0}|_{\kappa'=0} = 4.4\times 10^{-3}/0.7$ (right). The first corresponds, for example, to a realistic measurement rate and passive-cavity photon decay time of $\omega/2\pi = 300\,\text{Hz}$ and $1/|\gamma_{avg,0}|_{\kappa'=0} = 10\,\text{ns}$, respectively. The latter is more ambitious, corresponding to $\omega/2\pi = 1\,K\text{Hz}$ and $1/|\gamma_{avg,0}|_{\kappa'=0} = 1\,\mu\text{s}$. In both cases $\omega = \omega_h$ so that Eq. (37) is valid, as demonstrated by the fact that the orange and blue curves overlap in region II.

In Fig. 10(a) the frequency is sufficiently low that the system is in region I over the full range of detunings shown (the normalized frequency given above is small compared to the normalized decay rate, $|\gamma_{avg,0}|/|\gamma_{avg,0}|_{\kappa'=0} = 2\times 10^{-4}/0.7$). Note that the while the quasilinear theory (dotted green curve) predicts an enhancement in precision for certain detunings, the orange dashed curve shows that the precision is less than unity ($\varsigma \leq 1$) over most of the detuning range except for a small region around $\delta = 0$ where the enhancement is only slightly greater than unity. On the other hand, in Fig. 10(b) we have $\varsigma > 1$ in region II, but $\varsigma \leq 1$ in region I. The enhancement occurs because the frequency is nonzero, such that a crossing into region II is obtained at sufficiently low detunings. The same crossing does not occur in Fig. 10(a) owing to the subexceptional coupling and the low measurement frequency. However, in Fig. 10(c) the frequency is increased and again a crossing is observed into region II with a corresponding enhancement in precision. The effect of increasing the measurement frequency can also be seen in Fig. 10(d) where a large increase in the enhancement and bandwidth of region II is observed in comparison with Fig. 10(b). In addition, the enhancements in the figures on the right are larger than those predicted by the quasilinear prediction at small detuning.

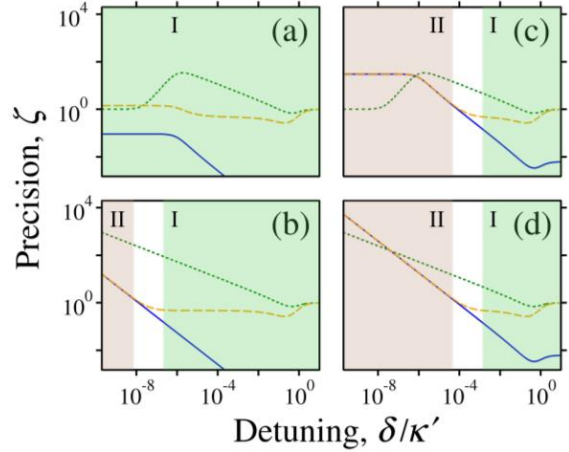

FIG. 10. Measurement precision for colored noise vs. detuning in the strong pumping limit, from the quasilinear prediction $S/K^{1/2}$ (green dotted curves) and from Eq. (36) in the white-noise limit (dashed orange curves) and in the random-walk white-noise limit (blue curves). The dashed orange curves are valid in the green-shaded region I, whereas the solid blue curves are valid in the red-shaded region II. The unshaded regions are transition regions. The frequencies are $\omega/|\gamma_{avg,0}|_{\kappa'=0} = 1.32\times 10^{-5}/0.7$ (left) and $4.4\times 10^{-3}/0.7$ (right), while the couplings are $\kappa'/\kappa^{EP} = 0.9999$ (top) and $\kappa'/\kappa^{EP} = 1$ (bottom). As in Fig. 6, $\gamma_1^n = \gamma_2^n = \kappa'$ and $\gamma_2^u = 0$.

## VIII. RECOVERING THE BEAT NOTE

While the analyses above illustrate the promise of conservative coupling, there is an important caveat: the beat frequency is zero in the regime that we are able to address with the linearization procedure, i.e., for $|\kappa' \sin 2\theta_0| < |\gamma_0/2|$. As we have pointed out previously [46] this single-lasing-eigenmode regime, which corresponds with the occurrence of LWG, is actually quite large, excluding only the case of exact unbroken PT-symmetry where $\delta = 0$ and $|\kappa'| \geq |\gamma_0/2|$. To understand why the beat frequency disappears, first note that it is given by $\dot{\theta}$ (just as it is for maximally dissipative coupling). By equating the threshold conditions from Eq. (22) and Eq. (26) we showed that the phase difference must be stationary, i.e, $\dot{\theta} = 0$, whenever the intensity is in steady state. Another way to understand this is that for any nonzero detuning, only one mode can lase with the other remaining below threshold. This is similar to the case of maximally dissipative coupling where one finds that $\dot{\theta} = 0$ within the deadband. As a result, the scale factor, and thus precision, are both zero.

There is an important distinction, however, that allows the beat frequency to be recovered in the case of conservative coupling, but not in the case of maximally dissipative coupling.

Although the relative frequency shift between the modes $\omega_+ - \omega_-$ is not observable (because only one mode lases), for conservative coupling there is still an absolute frequency shift for the mode that remains above threshold (there is no corresponding shift within the deadband because gain saturation ensures that $\gamma = 0$, see Eq. (7)). Therefore, we can recover the beat frequency simply by interfering the output with a reference frequency. There are essentially two ways to accomplish this: (*i*) by beating with an auxiliary mode such as an additional longitudinal (see Fig. 11 below), transverse, or polarization mode [32] in the same laser or in a separate laser, or (*ii*) by increasing the pumping to the extent that lasing can occur at both $\omega_+$ and $\omega_-$ simultaneously, i.e., the "two lasing eigenmodes" case [36].

These possibilities are shown in Table 1 for two different geometries: two coupled resonators (left column) and a single resonator whose coupler reverses the direction of the beam (right column). The second row elucidates the problem for the single-lasing-eigenmode case: the beat signal is formed by mixing two fields, $e_{1L}$ and $e_{2L}$, that copropagate inside the resonators with the same frequency, thus $\dot{\theta} = 0$. The output occurs either toward the left or right, but not both, and is arbitrarily chosen to be to the left in the figure. Similarly, the frequency is either $\omega_+$ or $\omega_-$, but not both, and is chosen to be $\omega_+$. The same situation occurs in the single resonator because when the system is unfolded it is apparent that the rightward-going wave for the single resonator is the same as $e_{2L}$ for the coupled resonator. Note the single resonator case is idealized because in practice a polarization difference is usually necessary to obtain the required loss difference.

For inhomogeneous gain, as the pumping increases it is possible that two eigenmodes lase. First, bidirectional operation becomes possible in the coupled resonator system as $\omega_+(\delta)$ and $\omega_+(-\delta)$ lase from two different output directions (third row). The negative sign on the detuning indicates that the sense of rotation is reversed for light at that output. Note that the frequency $\omega_+(-\delta)$ is identical to $\omega_-(\delta)$, but the threshold levels of the eigenmodes are different. Then, pumping further above threshold allows both $\omega_+(\delta)$ and $\omega_-(\delta)$ to lase in the single resonator system (fourth row). In this case the beat note appears even for a single output direction, but the mode intensities are unequal because one lases before the other, which reduces the beat note visibility. Thus, for inhomogeneous systems, both methods (*i*) and (*ii*) are possible, and the beat note may be recoverable without adding a second laser. On the other hand, when two lasing eigenmodes are not obtainable due to gain competition, insufficient pumping, or because the gain is homogeneous, the only possibility is to use method (*i*). Moreover, for homogeneous gain a second laser must be used. The top row in Table 1 shows a particular manifestation of method (*i*) that could be used for homogeneous broadening, where $\omega_+(\delta)$ and $\omega_+(-\delta)$ come from two distinct conservatively coupled systems, each operating in the single-lasing-eigenmode regime. Note that this situation is very similar to the bidirectional two-lasing-eigenmode case (third row), but with no mode competition.

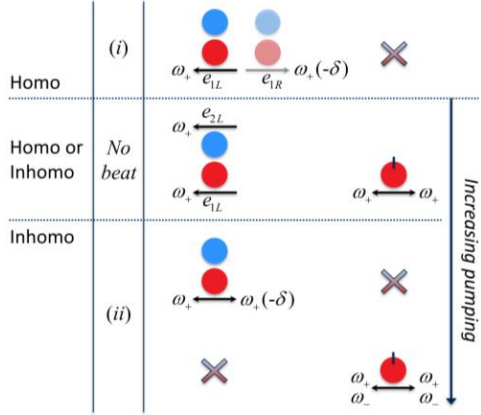

Table 1. Methods for recovering the beat note in conservatively coupled systems: two coupled resonators (left) and a single resonator with a coupler that reverses the direction of the beam (right). Homogeneous systems (top two rows) require interference with a second laser (shaded), whereas inhomogeneous systems (bottom three rows) can simply be pumped harder (pumping level is shown on the right). The 'x' indicates a method that is not possible for the particular system shown. The frequency at each output is indicated, assuming a rotation-induced detuning $\delta$, with the argument $-\delta$ indicating that the sense of rotation is reversed for light at that output. Laser emission does not occur from output directions that are not labelled. The coupler in the single resonator is represented by the vertical line. Where two frequencies are shown at a single output direction (bottom row), the top one lases first and with a higher intensity.

It is apparent, however, that homogeneously broadened single-lasing-eigenmode systems are not as interesting because they cannot achieve common mode. Yet our theoretical approach is only valid in the single-lasing-eigenmode regime, which applies for homogeneous or weakly-pumped inhomogeneous systems (second row). Indeed, it's important to point out that the solution of the stochastic nonlinear problem for the case of two lasing eigenmodes is very difficult near an EP and has not been solved for the case of conservative coupling. Nevertheless, we might gain some insight into the problem by a closer examination of the interference between the two single-lasing-eigenmode systems shown in the first row of Table 1. We assume there is no classical noise that would cause the phase of one laser to vary with respect to the other, and that the coupling between the two resonators is purely conservative. In addition, we assume the detunings of the two lasers are equal in magnitude and opposite in sign as a result of a rotation. The phase difference between the resonators for the leftward-going output, $e_{1L} = \mathcal{E}_{1L}\exp(i\phi_{1L})$, is denoted by $2\theta_L = \phi_{1L} - \phi_{2L}$. Similarly, $2\theta_R = \phi_{1R} - \phi_{2R}$ for the rightward-going output $e_{1R} = \mathcal{E}_{1R}\exp(i\phi_{1R})$. Each is in steady state such that there is no beat frequency ($\dot{\theta}_L = \dot{\theta}_R = 0$) for either laser output by itself. On the other hand, the phase difference between the two lasers is $2\theta_1 = \phi_{1L} - \phi_{1R}$, with a beat frequency given by

$$\dot{\theta}_1 = \delta/2 + (\kappa'/2)[1+\sin 2\chi]\cos^{-1}2\chi\cos 2\theta + f_{\theta 1}, \tag{38}$$

where $\theta = \theta_L$. Eq. (38) relates the beat frequency to the variables describing just one of the two (nearly) identical lasers. We have verified that this equation reproduces the beat frequencies shown in the inset of Fig. 5(b), as it should because $\omega_+(-\delta) = \omega_-(\delta)$. Note, however, that the noise in $\theta$ and $\chi$ will now be projected onto $\theta_1$, and that the cross correlations will no longer vanish, i.e., $\langle f_\theta(t)^* f_{\theta 1}(t')\rangle \neq 0$, $\langle f_\chi(t)^* f_{\theta 1}(t')\rangle \neq 0$, which can increase the excess-noise factor above that predicted in Section VI [59].

It is then worth considering to what extent the case of two lasing eigenmodes is represented by this simple superposition, which assumes no coupling, linear or nonlinear, between the two directions. In some situations, e.g., large detunings and no spatial overlap, the two lasing eigenmodes may hardly interact. Consider the case of a fast light gyroscope, for example, which uses two isotopes and a large bias to obtain bidirectional operation with no mode competition (as in Table 1, row 3) [28]. In this case, the superposition is valid. The two eigenmodes are uncoupled and may therefore be considered to come from two conservatively coupled single-mode lasers. For small detunings, on the other hand, the modes compete for gain, which could limit Petermann broadening (if both still lase). Thus, the simple superposition of the two single mode systems could represent a worst-case scenario as it does not take into account the cross saturation between the two modes. We should be careful, however, not to read too much into the above analysis, as the case of two lasing eigenmodes is still unsolved for conservative coupling.

Consider another example, illustrated in Fig. 11, where the beat note is recovered by taking advantage of the presence of an adjacent longitudinal mode (method (*i*)). In this experiment we employed a He-Ne laser tube with one flat window so that different output couplers could be selected. The length of the laser cavity was set sufficiently short (500.3 MHz free spectral range (FSR)) to support only two longitudinal modes (Fig. 11(b)). We coupled this to a second passive cavity that included a piezoelectric transducer (PZT) to change its length (Fig. 11(a)). The coupling was sufficiently weak that splitting of the laser mode never occurred ($|\kappa'|<|\gamma_0/2|$ when $\delta=0$, i.e., subexceptional regime). A variable absorber was placed in the second cavity to vary the loss and approach the EP (as well as to prevent feedback and chaos). We recorded the beat frequency between the longitudinal modes on a waterfall RF spectrum analyzer as the passive cavity modes were scanned across the laser modes (Fig. 11(b)), causing the laser FSR and beat frequency to oscillate (Fig. 11(c)). Each cycle of the oscillation corresponds to one FSR of the passive cavity (1.5 GHz). The oscillations increased as the absorption in the passive cavity decreased and its modes deepened (the passive cavity was undercoupled), demonstrating the increase in *S* as the EP was approached. At some point, however, the laser mode resonant with the passive cavity fell below threshold (see the dropouts in Fig. 11(c) when *a* = 0.2). A lower limit of $S=1.0008$ can be established by dividing the roughly 600 kHz deviation obtained at *a* = 0.2 by half the passive-cavity FSR. Higher gain would allow us to get closer to the EP and achieve larger values of *S*, nevertheless this simple experiment demonstrates the scale factor enhancement in the LWG (single lasing eigenmode) regime via recovery of the beat note. In this case the auxiliary lasing mode is far detuned from the coupled lasing mode. Still, mode competition cannot be entirely ignored because the standing wave geometry leads to some degree of spatial mode overlap. Therefore, while added quantum noise certainly enters the problem via the interference with the auxiliary mode, the situation is more complex than a simple superposition of the sort described above. More problematically, in this experiment classical noise is exacerbated because of the lack of common mode (the laser and passive cavities do not share a common path).

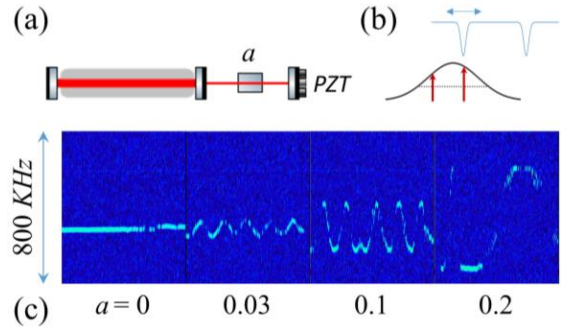

FIG. 11. (a) Beat frequency recovery by mixing with a second longitudinal mode in a laser coupled to a passive cavity. The coupling between the cavities never splits the lasing frequency,

thus in the absence of the second longitudinal mode there would be no beat frequency. (b) One of the passive cavity modes is repetitively scanned with a PZT across one of the laser modes, causing the beat frequency to oscillate. (c) The oscillation grows as absorption in the cavity decreases (*a* is the fraction of the electric field remaining after passing through the absorber) and the EP is approached.

In fact, common mode is generally difficult to obtain for conservative coupling even for the inhomogeneous two-lasing-eigenmode cases (Table 1, bottom two rows). Variations in coupling affect all the systems shown in Table 1, especially near the EP, and intensity variations can lead to a time-varying Kerr effect, but let us assume we can hold these factors constant. We will also assume there is no linear or nonlinear coupling between the two eigenmodes, so they can be treated as two independent single-lasing-eigenmode cases. Consider first the situation shown in the fourth row of Table 1. Assuming reciprocal and non-reciprocal (rotation-induced) inter-resonator detunings $\delta_r$ and $\delta_{nr}$, respectively, the beat frequency is given by

$$\omega_{beat} = \omega_+(\delta_a) - \omega_-(\delta_b)$$
$$= \frac{1}{2}[\Omega'(\delta_a) + \Omega'(\delta_b)] \xrightarrow{EP} |(\delta_r + \delta_{nr})\gamma/2|^{1/2}, \quad (39)$$

where we have assumed the outputs are from the same direction so that $\delta_a = \delta_b = \delta_r + \delta_{nr}$. Thus, common mode is only obtained when $\delta_r \ll \delta_{nr}$. In the single resonator geometry this is achieved since $\delta_r = 0$ when the optical paths are the same.

For the bidirectional situation (third row, left column) the beat frequency is $\omega_{beat} = \omega_+(\delta_a) - \omega_+(\delta_b)$, where $\delta_a = \delta_r + \delta_{nr}$ and $\delta_b = \delta_r - \delta_{nr}$ (only the nonreciprocal part is reversed). We then use the fact that $\omega_+(\delta_b) = \omega_-(-\delta_b)$ to write

$$\omega_{beat} = \omega_+(\delta_a) - \omega_-(-\delta_b) = \frac{1}{2}[\Omega'(\delta_a) + \Omega'(-\delta_b)]. \quad (40)$$

At the EP, this can be approximated as

$$\omega_{beat} = |\delta_{nr}\gamma/2|^{1/2}|(1+\delta_r/2\delta_{nr}) + (1-\delta_r/2\delta_{nr})|/2 \quad (41)$$

when $\delta_r \ll \delta_{nr}$. Common mode is again only obtained in this limit. Note, however, that the requirement for small $\delta_r$ is not as strict as in the unidirectional situation above, because of the subtraction. Indeed, Eq. (41) shows that, had the response been linear, the subtraction would have resulted in perfect common mode rejection irrespective of the size of the detuning. To better understand this, consider again the case of the dual-isotope fast-light gyroscope [28]. In the absence of a reciprocal detuning the two lasing eigenmodes are symmetric about the atomic absorption resonance. The presence of a reciprocal detuning, however, leads to an asymmetry such that the scale factor in one sense of rotation is different than the other, diminishing common mode. This asymmetry is minimized provided the detuning is sufficiently small that the gyroscope always operates in the regime where the dispersion associated with the atomic absorption can be considered linear. Note that in Eqs. (39-41) we have assumed the linear eigenvalues apply, but saturation can lead to an additional deviation from common mode conditions through the dependency $\gamma(\delta)$. To summarize, common mode can be approximated for the inhomogeneous two-lasing-eigenmode cases under the conditions outlined above. It requires either a shared path via a coupler that reverses the direction of the beam for the single resonator, or that the dispersion be linear for the bidirectional case.

### IX. COMPARISON WITH PREVIOUS EXPERIMENTAL RESULTS

In this section we review some previous experimental results that allow the relationship between *S* and *K* to be determined. In [41] the excess-noise factor *K* was measured in a quantum-noise-limited laser containing phase and loss anisotropy as the EP was approached.

This EP arises from maximally dissipative coupling and therefore corresponds to a lock-in edge. The loss anisotropy $\alpha$ was varied while the phase anisotropy (retardance) $\phi$ was held constant, verifying that the divergence of $K$ follows Eq. (3) in [41]. If in the same system $\phi$ were instead varied, while $\alpha$ was held constant, one could measure $S$. Lacking experimental data for this situation, we nonetheless have the functional form of the beat frequency (Eq. (1) in [41]) and can simply calculate $S$ by differentiating with respect to $\phi$ and normalizing to the case where $\alpha = 1$. In so doing we find the lock-in-edge EP at $\phi_c = \sin^{-1}[(1-\alpha^2)/(1+\alpha^2)]$, and that $|S| = K^{1/2}$ for all $|\phi| > |\phi_c|$, i.e., outside the lock-in region (corresponding to the green dashed curve in Fig. 3(b)). More recently, in [52] both $S$ and $K$ were measured outside the deadband of a Brillouin RLG, confirming the $|S| = K^{1/2}$ relationship in another system with maximally dissipative coupling. These results are understandable: as we have pointed out there can be no enhancement in precision for maximally dissipative coupling outside the deadband because $\gamma = 0$, and inside the deadband there is no beat note that can be recovered. Indeed, the very reason the linear theory has been shown to work so well in the cases analyzed thus far is because those cases involved dissipative coupling.

For the case of conservative coupling in lasers, the divergence of $S$ near the EP has been experimentally demonstrated [26, 28, 29, 36], but simultaneous measurements of the linewidth have yet to be reported. In [36], for example, an elegant non-Hermitian He-Ne ring laser gyroscope was constructed by creating a non-reciprocal loss difference between the two counterpropagating directions such that the system operates with two lasing eigenmodes, see Table 1 (fourth row, right column). As expected, this led to an increase in $S$ and decreased dead band. Common mode is largely preserved because a single ring is used, although the polarizations of the two directions are different. It is not clear, however, whether a saturation imbalance can be obtained in such single resonator geometries, nor whether this is even a requirement since the two-lasing-eigenmode case is still an unsolved problem. In contrast, fast-light gyroscopes [26, 28, 29] possess an intrinsic saturation imbalance between the laser cavity and dispersive medium. For the dual isotope gyro [28] the two lasing eigenmodes can be represented by two conservative single-lasing-eigenmode systems (that share the same path) since there is no mode competition, as in Eq. (38). However, the response is not linear, which should result in some loss of common mode. Thus, in the absence of linewidth measurements and lacking the solution to the two-lasing-eigenmode problem, it is not apriori obvious whether either sort of experiment can lead to enhanced precision.

In passive systems, on the other hand, there are several studies [27, 31] that report simultaneous measurement of the scale factor and linewidth in conservatively coupled systems. However, the extent to which our findings should apply to subthreshold systems is not entirely clear. It has been noted, for example, that excess noise relies on mode selection, i.e., amplification of spontaneously emitted photons, and does not apply to spontaneous processes far below threshold [60]. On the other hand, significant linewidth enhancements have been observed in passive-cavity systems [27, 32]. In these systems the uncertainty $\sigma$ depends on the cavity linewidth (proportionally as opposed to the square-root dependence on the Schawlow-Townes linewidth in lasers), the signal-to-noise at the detector, and on the shape of the resonance [25]. Each of these factors counteracts the increase in $s$ as the singularity is approached, limiting the enhancement in precision, similar to the role of $K$ in lasers. If the enhancement in uncertainty is defined as $\varepsilon = \sigma/\sigma_e$ where $\sigma_e$ is the uncertainty of the empty (or uncoupled) cavity, then the enhancement in precision is $|S|/\varepsilon$. It's not clear, however, whether in the quantum limit $\varepsilon$ and $K^{1/2}$ diverge equivalently or whether $|S|/\varepsilon$ can exceed unity. It has been shown that passive systems can be described as quasi-PT-symmetric through a simple decomposition of the Hamiltonian into PT-symmetric and lossy parts [12, 61]. On the other hand, it has been pointed out that the frequencies of the spectral extrema in subthreshold cavities do not generally coincide with the eigenvalue frequencies $\omega_\pm$ because the presence of the input can strongly influence the resulting spectrum, such that singularity (which is due to spectral splitting) does not occur at the EP [46]. As a result, not even $S$ diverges at the same rate as it does in lasing systems.

Whereas our approach is semiclassical, several recent studies have taken a fully quantum approach, finding lower bounds for the error in the estimation of the perturbation (detuning in our case) by treating these CR systems as linear Markovian open quantum systems via quantum Langevin equations [62-64]. These studies have been limited to subthreshold systems with linear matrices. In [64] the Cramér-Rao bound was found to be significantly lower at the EP when operating close to the laser threshold; however, it has been pointed out that the assumption of linearity may be problematic in this regime [63]. In [62] and [63] no advantage was found from operating near an EP. These studies did not examine the near-threshold regime but appear to rule out purely passive systems and systems with lower levels of amplification.

In [31] the frequency shift and linewidth difference were measured in a passive micro-cavity perturbed by nanoscale scatterers to approach the EP. Although the uncertainty was not explicitly determined, Langbein [65] has carefully analyzed this data and shown that the precision in fact decreased as the EP is approached, such that $|S|/\varepsilon$ was always less than unity. In [27] $S$ and $\varepsilon$ were explicitly measured in a passive-cavity due to the presence of a fast-light medium as the critical anomalous dispersion was approached. This data also showed $|S|/\varepsilon<1$. Neither of these experiments were performed near the quantum limit, however. Nor did they involve amplification. Therefore, they do not resolve whether an EP-related enhancement in precision is possible in either passive or near-threshold CR systems.

## X. SUMMARY AND CONCLUSIONS

Our results explain why increased measurement precision by the use of EPs is not straightforward to demonstrate. In the linear regime, EPs are rotationally invariant and do not boost precision, regardless of the type of EP. Indeed, the enhancement in precision is never greater than unity for any set of parameters. Instead, in the vicinity of an EP a hole of reduced precision opens up where the precision drops precipitously to zero within regions of deadband or unbroken PT-symmetry. This behavior is universal, with the hole appearing regardless of the type of EP. Outside of these zero-sensitivity regions the precision approaches its maximum value of $|S|/K^{1/2}=1$ at the EP (the Petermann limit). EPs, therefore, represent discontinuous transitions between these two regions.

To describe systems pumped above threshold we took first a quasilinear approach. We found that the Petermann limit also applies to maximally dissipative EP systems, which explains why experiments in conventional laser gyroscopes have found no increase in precision at the deadband edge EP [35]. In addition, this approach revealed that (at least for the single-lasing-eigenmode case) the Petermann limit applies when the coupling is conservative and there is no saturation imbalance between the resonators. The presence of a saturation imbalance, on the other hand, can break the rotational symmetry of EPs, leading to an additional enhancement in $S$ for conservative coupling beyond that usually expected near an EP. We then took a more intrinsically nonlinear approach by linearizing the coupled equations around their stable stationary solutions, and found that the excess noise factors in phase and especially intensity can be even lower in these conservative CRs than predicted by the quasi-linear theory, owing to the coupling between the noise fluctuations and the fact that the threshold conditions are different for the coupled and uncoupled systems (which causes the steady state gain coefficients and conventional linewidths to be different). We verified these results through numerical solutions of the coupled equations with Langevin noise present. In addition, we found that because the coloring of the noise is described by different characteristic decay rates for the two systems, an enhancement in precision beyond the Petermann limit can occur when the measurement frequency is sufficiently high that white frequency noise dominates in the coupled system while white phase noise dominates in the uncoupled system.

On the other hand, we showed that for our system of equations only one mode lases. The beat note can be recovered by interference with an auxiliary mode, or by pumping into the regime of two lasing eigenmodes. In both case there are consequences for the quantum and classical noise that depend on the details of the recovery method. More specifically, our prediction of an enhancement in quantum-limited precision coincides with an ideal situation (interference with a separate laser with a much narrower linewidth [59]) that cannot be common mode. Deviating from this ideal situation leads either to increased quantum noise (see Eq. (38)) or into the two-lasing-eigenmode regime, which is an unsolved problem. Thus, while our results clarify key ingredients for enhancing the fundamental precision in EP systems (conservative coupling, LWG, a saturation imbalance between the resonators, operation sufficiently close to the EP and at sufficiently high measurement frequencies, and recovery of the beat note with minimum added noise), it is not clear whether practical experiments can be devised that meet all of these concurrent requirements. Moreover, it is not a priori obvious whether all these requirements will apply in the two-lasing-eigenmode case.

In addition, we have not explicitly evaluated the effect of deviations from these ideal conditions such as the effect of small amounts of dissipation or nonreciprocity [62] in the coupling, other accompanying nonlinearities, or non-vanishing cross-correlations between noise sources (which depends on the particular beat-note recovery method). Furthermore, the linearization procedure becomes problematic directly at the EP. Note in Eq. (28a) that when $\gamma_{avg,0} = 0$, the characteristic time to return to steady-state diverges and the nonlinear terms in the analysis can no longer be ignored, which has implications for the stability of the steady state as well as for the calculation of the noise spectrum. This is also the case for the previous results obtained for the maximally dissipative EP at the conventional gyro deadband edge [58].

More work also needs to be done to understand the role excess noise plays in the bad cavity limit, where the laser linewidth is determined more by the gain bandwidth than by the cavity linewidth and can be substantially reduced as a result [66]. Certainly, the Petermann broadening can never be larger than the gain bandwidth in any laser. The authors have observed a similar clamping of the broadening in passive non-Hermitian cavities [25, 32], and while this effect clearly increases the scale factor to linewidth ratio, it's unclear whether it leads to increased precision. Finally, solution of the case of two lasing eigenmodes will ultimately be necessary for evaluating the Petermann broadening in inhomogeneous systems, in particular where cross saturation plays a role. Note, for example, that the predicted enhancement in precision in Fig. 10 occurs at small $\delta$, precisely where mode competition will be most egregious. Indeed, mode competition will rule out the bidirectional operation shown in Table 1 in a large number of systems.

An additional benefit of this work may accrue from the equivalency between the critical anomalous dispersion in fast-light systems and EPs in coupled resonators [32, 67]. While the increased linewidth in lasers with nonorthogonal eigenmodes was demonstrated long ago, it is only recently that measurements of increased scale factor have begun to appear. These studies often tacitly ignore the change in linewidth. For lasers containing fast-light media this has, at least in part, been due to the lack of theoretical predictions regarding their fundamental linewidth. Characterizing fast-light systems by an excess noise factor may provide an alternative way to estimate their linewidth [68].

## XI. ACKNOWLEDGMENTS

D. D. Smith and H. Chang were supported internally through the Technology Investment Program of the Marshall Space Flight Center (MSFC). E. Mikhailov was supported by the NASA MSFC Faculty Fellow Program. S. M. Shahriar was supported by AFOSR Grant # FA9550-18-1-0401 and NASA Grant # 80NSSC21C0172. The authors acknowledge helpful discussions with F. Bretenaker, K. Myneni, J. C. Diels, and L. Horstman.

classical noise due to lack of common mode, the quantum limited precision for the beat frequency is the same as that predicted for the zero-beat-frequency single-lasing-eigenmode case in section VII.